\documentclass[10pt]{article}

\usepackage{grffile}
\usepackage[margin=.9in]{geometry}
\usepackage{amssymb}
\usepackage{amsmath}
\usepackage[dvips]{epsfig}
\usepackage[small]{caption}
\usepackage{graphicx}
\usepackage[all]{xy}

\usepackage[hidelinks]{hyperref}
\usepackage{caption}
\usepackage{subcaption}
\usepackage{colortbl}
\usepackage{mathrsfs}
\usepackage{tikz}
\usetikzlibrary{positioning}
\usepackage{enumitem}

\newtheorem{Lemma}{Lemma}[section]
\newtheorem{Theorem}{Theorem}
\newtheorem{Proposition}[Lemma]{Proposition}
\newtheorem{Corollary}[Lemma]{Corollary}
\newtheorem{Remark}[Lemma]{Remark}

\newenvironment{Proof}%
 {\begin{trivlist} \item[]{\bf Proof. }}%
 {\hspace*{\fill}$\rule{.4\baselineskip}{.4\baselineskip}$\end{trivlist}}

 {\begin{trivlist}\item[]\textbf{Acknowledgments.}}{\end{trivlist}}

\makeatletter\@addtoreset{figure}{section}\makeatother

\renewcommand{\theTheorem}{\arabic{Theorem}}

\makeatletter \@addtoreset{equation}{section} \makeatother

\newcommand{\R}{\mathbb{R}}

\newcommand{\rmO}{\mathrm{O}}

\renewcommand{\leq}{\leqslant}
\renewcommand{\geq}{\geqslant}

\def\Xint#1{\mathchoice
   {\XXint\displaystyle\textstyle{#1}}%
   {\XXint\textstyle\scriptstyle{#1}}%
   {\XXint\scriptstyle\scriptscriptstyle{#1}}%
   {\XXint\scriptscriptstyle\scriptscriptstyle{#1}}%
   \!\int}
\def\XXint#1#2#3{{\setbox0=\hbox{$#1{#2#3}{\int}$}
     \vcenter{\hbox{$#2#3$}}\kern-.5\wd0}}

\def\dashint{\Xint-}

\begin{document}

\title{Phase separation patterns from directional quenching} 

\author{ Rafael Monteiro  and  Arnd Scheel  \\[2ex]
\textit{\footnotesize University of Minnesota, School of Mathematics,  
206 Church St. S.E., Minneapolis, MN 55455, USA}} 

\date{\small \today} 

\maketitle

\begin{abstract}
We study the effect of directional quenching on patterns formed in simple bistable systems such as the Allen-Cahn and the Cahn-Hilliard equation on the plane. We model directional quenching as an externally triggered change in system parameters, changing the system from monostable to bistable across an interface. We are then interested in patterns forming in the bistable region, in particular as the trigger progresses and increases the bistable region. We find existence and non-existence results of single interfaces and striped patterns.

\vspace{\baselineskip}

\smallskip
\noindent \textbf{Keywords.} Phase separation, directional quenching, Allen-Cahn, Cahn-Hilliard.
\end{abstract}

\section{Introduction}

We are interested in phase separation patterns arising when systems parameters are varied across an interface moving with constant speed, such that the system undergoes a phase separation process in the wake of the interface. As a simplest model for phase separation, we start with a bistable, double-well energy and a surface energy term, 
\[
\mathscr{E}[u]=\int_{x,y} \left(\frac{1}{2}|\nabla u|^2 + \frac{1}{4}(\mu-u^2)^2\right) dx dy,
\]
with preferred minima $u=\pm \sqrt{\mu}$. In the simplest case, we think of $\mu=\mu(x)=-\mathrm{sign}\,(x)$, rendering the medium bistable in the half plane $x<0$, and monostable in $x>0$.  More interestingly, we are interested in the dynamic question, where $\mu=-\mathrm{sign}\,(x-ct)$, $c>0$, that is, the medium is bistable in the growing region $\{(x,y)|\,x<ct\}$. We refer to this setting as directional quenching in the $x$-direction with speed $c$. Our focus here is on slow quenching, $0\leq c\ll 1$. We note that the case $c<0$ is mathematically perfectly valid but possibly not as interesting as $c>0$: phase separation patterns could be created for $c>0$ rather than annihilated at the interface for $c<0$. Indeed, critical points of the energy for $\mu\equiv 1$ include a plethora of phase separation patterns \cite{Fife_models}, that is, solutions with nodal lines separating regions with $u>0$ from regions where $u<0$, including simple straight interfaces $u=u(x)\to \pm 1$ as $x\to \pm\infty$, and 
periodic stripes $u=\bar{u}(x;\kappa)$,
\begin{equation}\label{e:per}
\bar{u}''+\bar{u}-\bar{u}^3=0, \qquad \bar{u}(x;\kappa)=-\bar{u}(x+\kappa;\kappa)=-\bar{u}(-x;\kappa)\not\equiv 0,\qquad \mbox{ for } x\in\R,
\end{equation}
with half-periods  $\pi<\kappa<\infty$, and normalization $\bar{u}'(0)>0$.
  
In order to study the dynamic setup, where $c>0$, we focus on two gradient flows associated with the energy $\mathscr{E}$, the $L^2$-gradient flow with associated Allen-Cahn equation (AC), and the $H^{-1}$-gradient flow with associated Cahn-Hilliard (CH) equation. In a comoving frame $\tilde{x}=x-ct$, those equations read
\begin{equation}\label{Allen-Cahn}
u_t  = \Delta u + \mu(x) u  - u^3 + c u_x \qquad \qquad  \mbox{(Allen-Cahn)},
\end{equation}
and
\begin{equation}\label{Cahn-Hilliard}
u_t  = -\Delta(\Delta u + \mu(x) u  - u^3) + c u_x \qquad  \mbox{(Cahn-Hilliard)},
\end{equation}
where $\mu(x)=-\mathrm{sign}\,(x)$, $c\geq 0$, $(x,y)\in\R^2$, subscripts denote partial derivatives, and we dropped the tilde for ease of notation. Our focus will be on stationary solutions, $u_t=0$, and we will only briefly comment on relevant solutions with nontrivial time dependence. Throughout, we will be thinking of $c>0$ small as a perturbation parameter, starting from the zero speed case. 

We are not aware of a systematic study of directional quenching processes in the mathematical literature. Our work here is motivated to a large extent by phase separation processes in recurrent precipitation \cite{liese,Droz2000,thomas2013helices,lagziprl}, studies of patterning in Langmuir–Blodgett transfer \cite{langmuir2,langmuir,thiele}, and numerical studies in \cite{foard2012survey}. 
\begin{figure}
\includegraphics[height=1.8in]{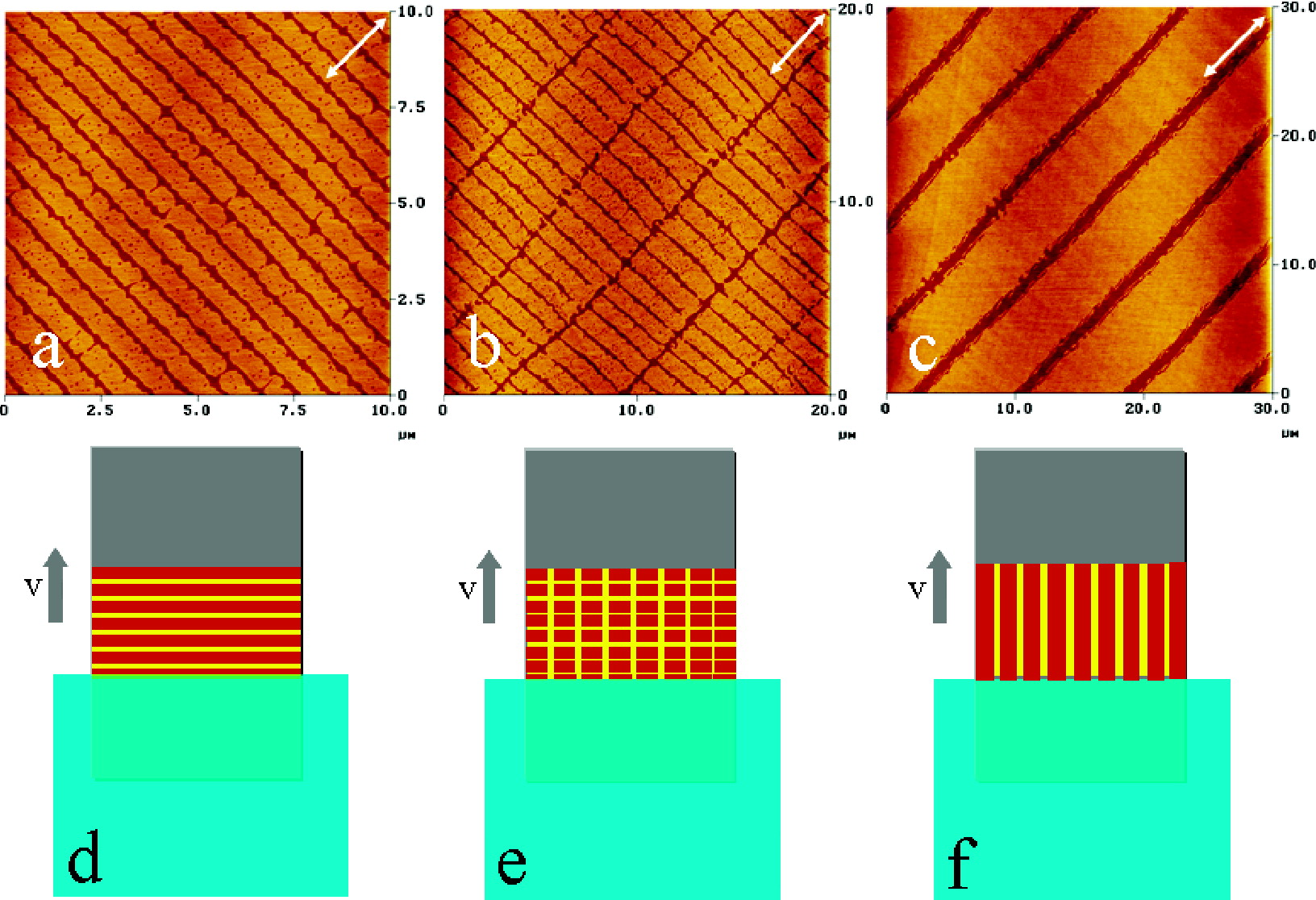}\hfill \includegraphics[height=1.8in]{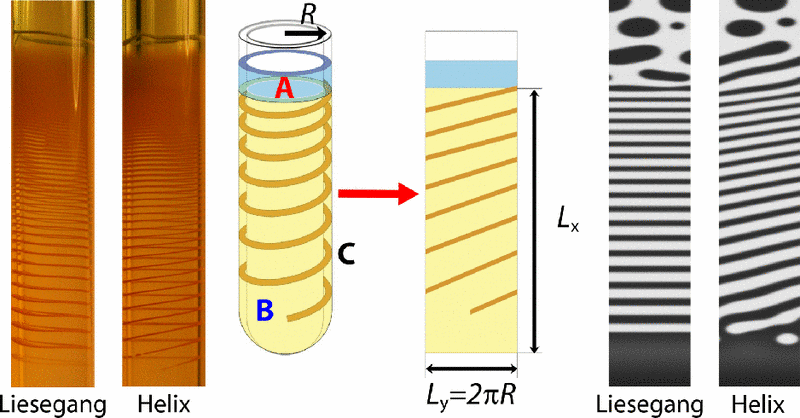}
\caption{ Shape and alignment of patterns arising in Langmuir-Blodgett transfer of  a homogeneous $L$-$\alpha$--dipalmitoylphosphatidylcholine Langmuir transfer; reproduced with permission from \cite{langmuir2}. Copyright 2007, ACS."
monolayer (right). Liesegang rings and helices formed through recurrent precipiation in tube-in-tube experiments  $Cu^{2+}(aq)+CrO^{2−}_4(aq)\to CuCrO_4(s)$ in 1\% agarose gel, schematic of relation to 2d-patterning, and numerical simuilations; reproduced with permission from \cite{lagziprl}, Copyright 2013, APS.
\label{f:exp}}
\end{figure}

Both experimentally and numerically, a plethora of patterns can be observed depending on initial conditions and parameter values. One particular question of interest there is the orientation of interfaces: depending on system parameters and initial conditions interfaces parallel, perpendicular, as well as slanted relative to the quenching boundary $\{x=0\}$ are observed. Our results can roughly be understood as establishing the existence of stripes perpendicular to the interface and ruling out slanted stripes. Stripes parallel to the interface were found in an asymptotic analysis in \cite{krekhov2009formation} for the Cahn-Hilliard equation. We rule out 
the creation of  stripes parallel to the interface in the Allen-Cahn equation. 

In the case of zero speed, solutions to the Cahn-Hilliard equation solve 
\begin{equation}\label{ach}
  \Delta u + \mu(x) u  - u^3 = \nu,
\end{equation}
 where $\nu$ is usually referred to as the chemical potential. If we require zero mass, $u\to 0$ as $x\to \infty$, we find $\nu=0$ and we recover the Allen-Cahn problem at $c=0$
 \begin{equation}\label{ach0}
  \Delta u + \mu(x) u  - u^3 = 0,
\end{equation}
 As a consequence, much of the present work treats both cases simultaneously. We note here that the unbalanced cases, $\nu\neq 0$, as well as more generally unbalanced or even non-odd nonlinearities pose significant obstacles to the analysis here and likely give rise to different phenomena. 
 
 We remark here that somewhat related problems arise in the context of ecology, where a change of stability of the trivial state encodes a spatial boundary to the habitat of a species, that is, to the region, where small populations can grow and spread. Much recent work has focused on the effect of shifting habitats due to say climate change, and the question whether species can follow the spatial shift; see for instance \cite{ber,giletti2,giletti1,bingtuan,vo} and the references therein. As we explained above, the main thrust of the present work is towards the characterization of patterns in the wake of such shifting boundaries, slightly different from the major questions arising in the context of ecology. 

In the remainder of the introduction, we first characterize more precisely possible morphologies in the wake of the quenching process, then state our main results for speeds $c=0$ and then $c>0$, briefly discuss methods employed in the proofs, and briefly discuss some of the many open questions. 
 
\subsection{Stripe mopholologies in the wake.}

Depending on the pattern in the wake of the quenching process, we distinguish four cases of interest to us here. Our terminology refers to an orientation where the quenching process progresses ``horizontally'', in the $x$-direction. 

\begin{figure}[htb]
    \centering
    \begin{subfigure}[b]{0.45\textwidth}
          \centering
        \includegraphics[height = 0.4in]{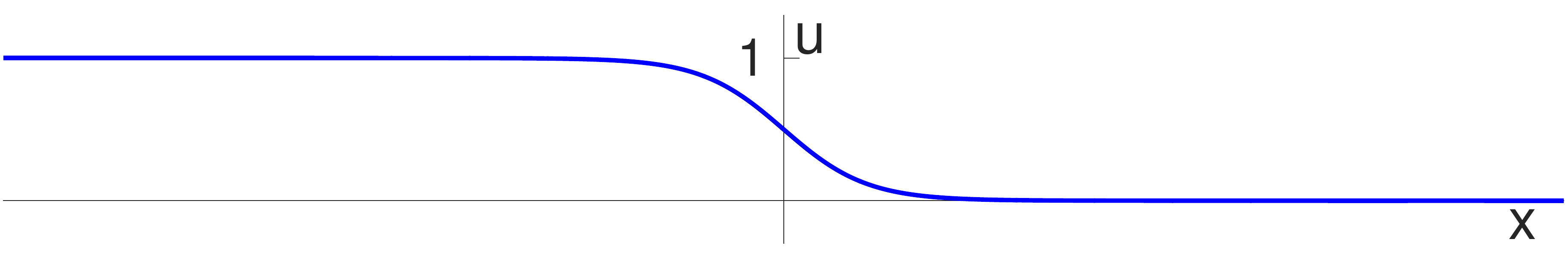}
    \end{subfigure}
    ~ 
    \begin{subfigure}[b]{0.45\textwidth}
    \centering
    \includegraphics[width=\textwidth,height=0.4in]{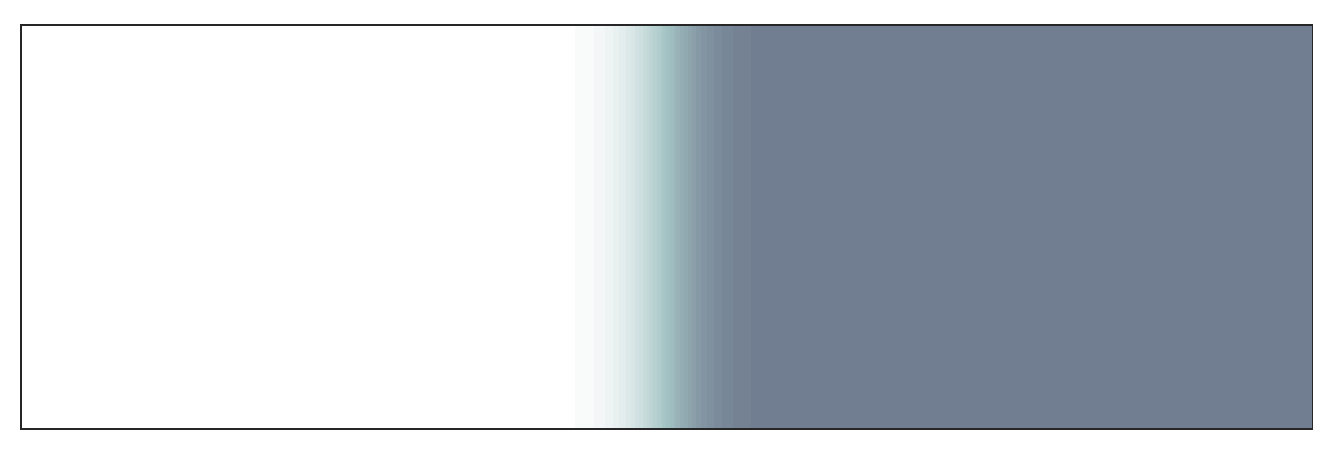}
        \end{subfigure}
    \caption{Pure phase selection $1 \leadsto 0$, solution $u(x)$ (left) and contour plot for $(x,y)\in\R^2$ (right). \label{Figure_1-1_leadsto_0}}
\end{figure}

\paragraph{Pure phase selection --- \textbf{$1\leadsto 0	$} fronts}
The simplest and least interesting case is when the quenching process generates a pure phase, that is, it does not generate  interfaces between regions where $u>0$ and $u<0$, respectively. 
Such solutions can of course be found in one space dimension, $x\in\R$, requiring $\lim_{x \to -\infty} u(x) = + 1$ and $\lim_{x \to \infty} u(x) = 0$; see Figure \ref{Figure_1-1_leadsto_0}. 
%

We will see that such solutions exist at $c\gtrsim 0$ for AC but only at $c=0$ for  CH.

\begin{figure}[htb]
    \centering
    \begin{subfigure}[b]{0.45\textwidth}
          \centering
          \includegraphics[height=0.4in]{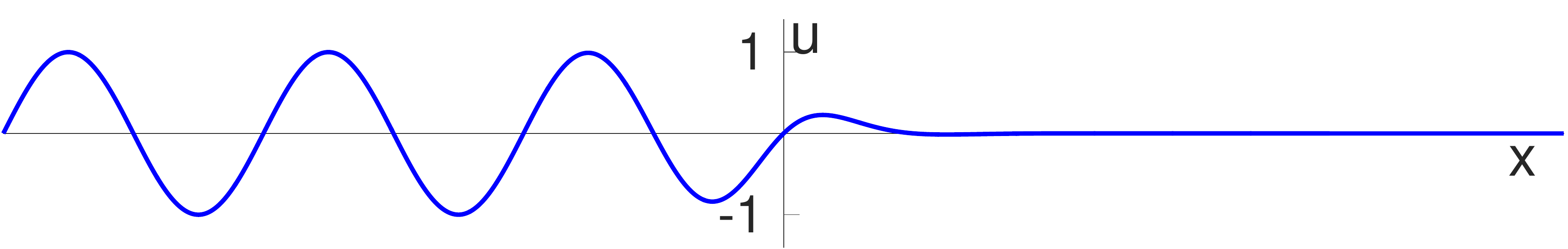}
    \end{subfigure}
    ~ 
    \begin{subfigure}[b]{0.45\textwidth}
    \centering
    \includegraphics[width=\textwidth,height=0.4in]{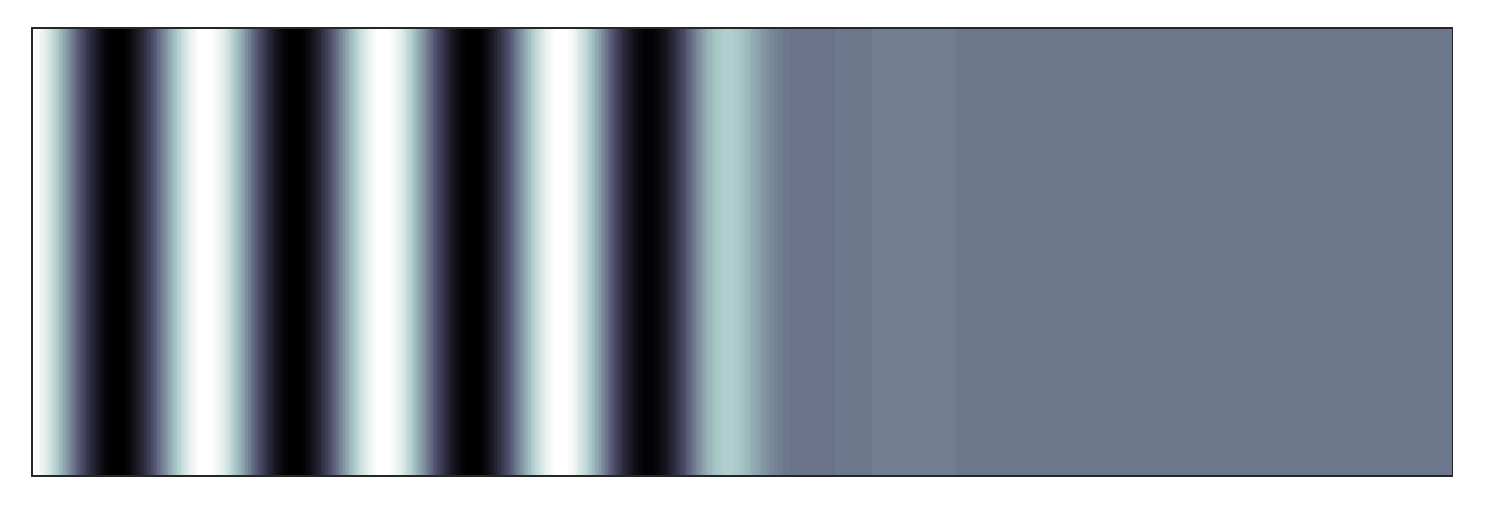}
        \end{subfigure}
    \caption{Vertical stripes $\mathcal{V}$, solution $u(x)$ (left) and contour plot  for $(x,y)\in\R^2$ (right). \label{Figure_1-vertical}}
\end{figure}

\paragraph{Vertical stripes ($\mathcal{V}$).}
An essentially one-dimensional problem is the formation of vertical stripes in the wake of the quenching process; see Figure \ref{Figure_1-vertical}. We look for solutions with $\lim_{x \to \infty} u(x) = 0$, $\lim_{x \to -\infty} |u(x)-\bar{u}(x;\kappa)| = 0$ for some $\kappa>0$.

We will see that such solutions exist for $c=0$. For $c>0$, such solutions would necessarily be time-dependent, since the asymptotic pattern $\bar{u}$ is time-periodic. It turns out that such time-periodic solutions do not exist in AC. For CH, a matched asymptotics analysis in \cite{krekhov2009formation}, supported by numerical simulations, shows that such time-periodic solutions exist in CH for small speed. Moreover, the wavenumber is computed as  a function of the quenching speed $k=k(c)=\rmO(c)$ as $c\to 0$. Such patterns can be viewed as a prototype of the horizontally banded stripes in Liesegang band generation \cite{Droz2000}. 

\begin{figure}[htb]
    \centering
    \begin{subfigure}[b]{0.45\textwidth}
          \centering
          \includegraphics[width=\textwidth,height=0.4in]{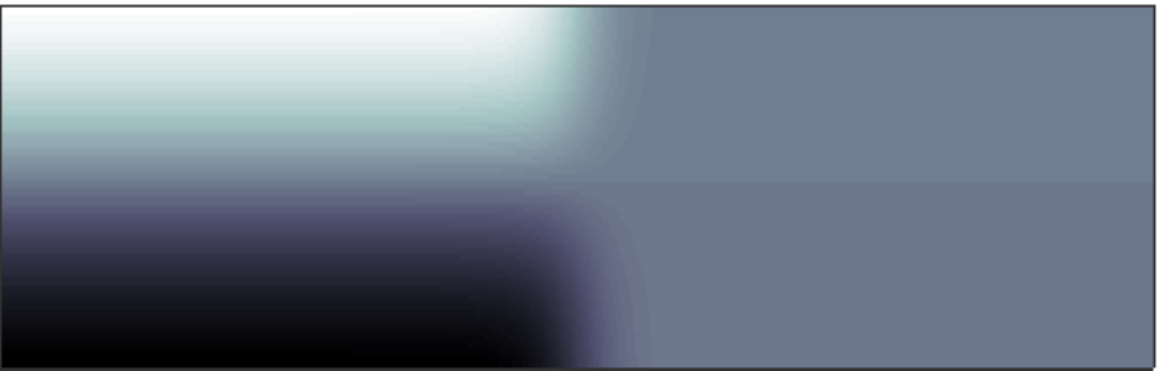}\\[0.2in]
          \includegraphics[width=\textwidth,height=0.4in]{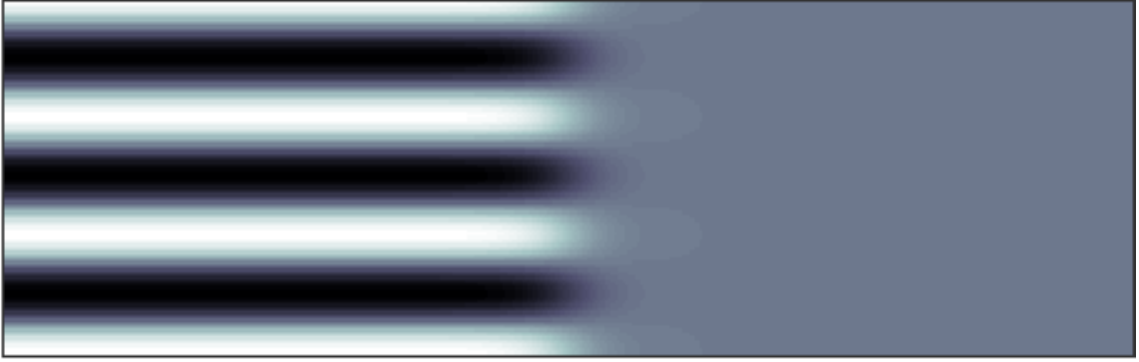}
              \end{subfigure}
    ~ 
    \begin{subfigure}[b]{0.45\textwidth}
    \centering
    \includegraphics[width=\textwidth,height=0.4in]{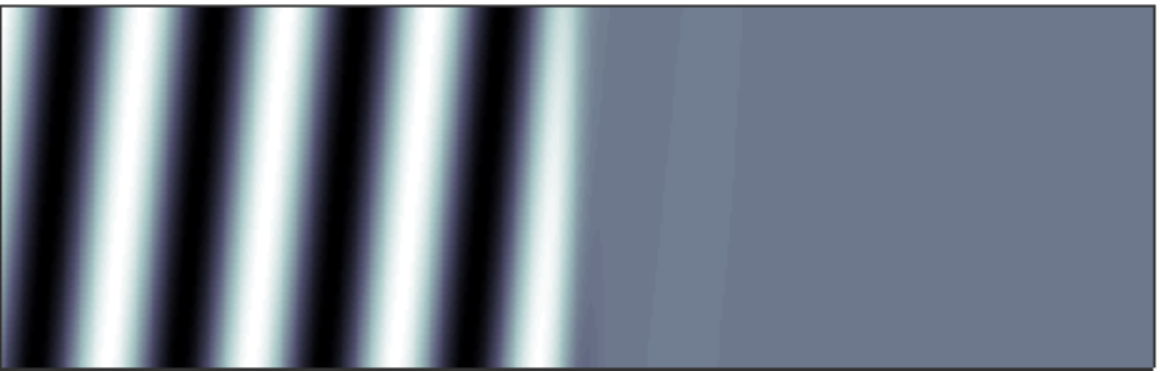}\\[0.2in]
    \includegraphics[width=\textwidth,height=0.4in]{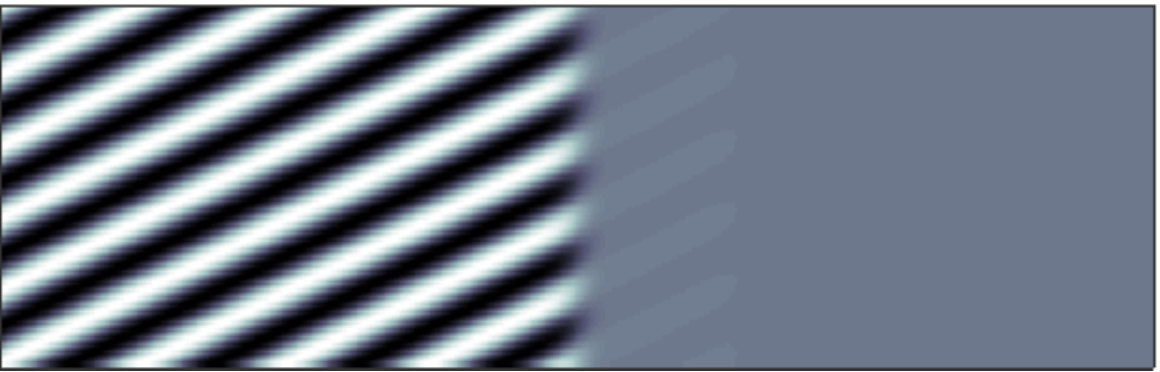}
        \end{subfigure}
    \caption{Horizontal patterns,  $\mathcal{H_\infty}$ (top left) and  $\mathcal{H}$ (bottom left)  and  oblique stripes with small and large angles relative to $x=0$, respectively (right). \label{Figure_1-horizontal}}
\end{figure}

\paragraph{Horizontal ($\mathcal{H}$,$\mathcal{H_\infty}$) and oblique stripes ($\mathcal{O}$).}
True two-dimensional solutions arise when the periodic patterns are not vertically oriented; see Figure \ref{Figure_1-horizontal}. Solutions in this case have asymptotics $\lim_{x\to -\infty}|u(x,y)-\bar{u}(\cos(\phi) x + \sin(\phi) y;\kappa)|\to 0$, where $\phi\in [0,\pi/2]$. Horizontal stripes correspond to $\phi=\pi/2$. A particularly interesting limiting case is the creation of one interface, $\lim_{x\to -\infty}|u(x,y)-\tanh(y/\sqrt{2})|\to 0$.

%
%

We will see that oblique stripes do not exist at $c=0$, horizontal stripes exist for $c=0$ and $c\gtrsim 0$, in both AC and CH. Single interfaces exist at $c=0$, and for $c\gtrsim 0$ in AC but not in CH. Curiously, oblique stripes are observed for finite speed \cite{thomas2013helices} and their existence region remains a challenging problem in this area.

We remark that we do not discuss the possibility of cellular patterns emerging in the wake of the quenching process, a phenomenon of interest and observed in \cite{foard2012survey}, but beyond the reach of our methods here. 

\paragraph{Nomenclatura and brief summary.} We refer to the existence of the pure phase fronts as the $1\leadsto 0$-problem, the creation of vertical, horizontal, and oblique stripes as the $\mathcal{V}$-, $\mathcal{H}$-, and $\mathcal{O}$-problem, and the creation of a single interface as the $\mathcal{H}_\infty$-problem. A grasshoppers guide to our main  results is contained in the following table.

\begin{center}
\begin{tabular}{cccccc  }
& $1\leadsto 0$ &  $\mathcal{V}$ & $\mathcal{H}_\infty$ & $\mathcal{H}$ & $\mathcal{O}$ \\[0.03in]
\hline\\
\begin{minipage}{2.4cm}
\begin{center}
$c=0$\\[0.03in]
AC/CH
\end{center}
\end{minipage}
& 
\begin{minipage}{2.4cm}
\begin{center}
yes\\[0.03in]
Prop \ref{main_theorem:c_zero:1_leadsto_0}
\end{center}
\end{minipage}
&
\begin{minipage}{2.4cm}
\begin{center}
yes\\[0.03in]
Prop \ref{main_theorem:c_zero:1_leadsto_0}
\end{center}
\end{minipage}
&
\begin{minipage}{2.4cm}
\begin{center}
yes\\[0.03in]
Thm \ref{main_theorem:c_zero:H_infty}
\end{center}
\end{minipage}
&
\begin{minipage}{2.4cm}
\begin{center}
yes\\[0.03in]
Thm  \ref{main_theorem:c_zero:H_infty}
\end{center}
\end{minipage}
&
\begin{minipage}{2.4cm}
\begin{center}
no\\[0.03in]
Rem \ref{r:onon}
\end{center}
\end{minipage}
\\[0.03in]
\\
 \hline\\
\begin{minipage}{2.4cm}
\begin{center}
$c\gtrsim0$\\[0.03in]
AC
\end{center}
\end{minipage}
& 
\begin{minipage}{2.4cm}
\begin{center}
yes\\[0.03in]
Prop \ref{t:acc}
\end{center}
\end{minipage}
&
\begin{minipage}{2.4cm}
\begin{center}
no\\[0.03in]
Rem \ref{p:nodal}
\end{center}
\end{minipage}
&
\begin{minipage}{2.4cm}
\begin{center}
yes\\[0.03in]
Prop \ref{t:acc}
\end{center}
\end{minipage}
&
\begin{minipage}{2.4cm}
\begin{center}
yes\\[0.03in]
Prop \ref{t:acc}
\end{center}
\end{minipage}
&
\begin{minipage}{2.4cm}
\begin{center}
no\\[0.03in]
Rem \ref{r:noo}
\end{center}
\end{minipage}
\\[0.03in]
\\
 \hline\\
\begin{minipage}{2.4cm}
\begin{center}
$c\gtrsim0$\\[0.03in]
CH
\end{center}
\end{minipage}
& 
\begin{minipage}{2.4cm}
\begin{center}
no\\[0.03in]
Rem \ref{r:noch}
\end{center}
\end{minipage}
&
\begin{minipage}{2.4cm}
\begin{center}
(yes)\\[0.03in]
\cite{krekhov2009formation}
\end{center}
\end{minipage}
&
\begin{minipage}{2.4cm}
\begin{center}
no\\[0.03in]
Rem \ref{r:noch}
\end{center}
\end{minipage}
&
\begin{minipage}{2.4cm}
\begin{center}
yes\\[0.03in]
Prop \ref{t:chc}
\end{center}
\end{minipage}
&
\begin{minipage}{2.4cm}
\begin{center}
not known
\end{center}
\end{minipage}
\\[0.03in]
\\

 \hline
\end{tabular}
\end{center}

\subsection{Main results --- zero speed} 

We will state existence and nonexistence results for zero speeds, that is, we will discuss existence and nonexistence of solutions to \eqref{ach0} for $\nu=0$, with prescribed asymptotics at spatial infinity. Since the coefficients of this elliptic equation are discontinuous at the line $x=0$, solutions will be classical only for $x\neq 0$, and weak solutions, H\"older continuous together with their first derivatives across $x=0$. We will in the following results simply refer to such functions as \emph{solutions} and emphasize the behavior at infinity. 

Our first result concerns the essentially one-dimensional case. Recall the definition of the one-dimensional periodic solutions $\bar{u}(x;\kappa)$ with period $2\kappa$ from \eqref{e:per}.
\begin{Proposition}[$1 \leadsto 0$, $\mathcal{V}$]\label{main_theorem:c_zero:1_leadsto_0} There exists a unique family of solutions $\theta(x;\kappa)$, $\kappa\in (\pi,\infty]$,  to \eqref{ach0} such that we have  $\lim_{x\to\infty}\theta(x;\kappa)=0$, and,
\begin{align*}
\kappa=\infty:\qquad\qquad &  \lim_{x\to -\infty}\left( \theta(x;\infty)-1\right)=0;\\
\kappa<\infty:\qquad \qquad & \lim_{x\to -\infty} \left(\theta(x;\kappa)-\bar{u}(x-\xi(\kappa);\kappa)\right)=0  \mbox{ for some smooth function } \xi(\kappa).
\end{align*}
Moreover, $\theta(x;\infty)\in (0,1)$ and $\theta(x;\kappa)\in (-1,1)$, for all $x\in\R,\ \kappa\leq \infty$. 
\end{Proposition}
Since this result is concerned with an ordinary differential equation, the proof follows by simple phase plane analysis. We shall give some details in this direction in Section \ref{section:1_leadsto_0} but also give a longer proof that relies on comparison principles and builds the basis for the proof of the higher-dimensional results that we shall state next. 
\begin{Theorem}[$\mathcal{H}$, $\mathcal{H}_\infty$]\label{main_theorem:c_zero:H_infty} There exists a family of solutions $\Xi(x,y;\kappa)$, , $\kappa\in (\pi,\infty]$,  to \eqref{ach0} such that we have  $\lim_{x\to\infty}\Xi(x,y;\kappa)=0$, and
\begin{align*}
\kappa=\infty:\qquad\qquad &  \lim_{x\to -\infty}\left( \Xi(x,y;\infty)-\tanh(y/\sqrt{2})\right)=0;\\
\kappa<\infty:\qquad \qquad & \lim_{x\to -\infty} \left(\Xi(x,y;\kappa)-\bar{u}(y;\kappa)\right)=0.
\end{align*}
Moreover, the convergence is exponential, uniformly in $y$. The solutions have symmetries
\begin{align*}
\kappa=\infty:\qquad\qquad &  \Xi(x,y;\infty)=-\Xi(x,-y;\infty);\\
\kappa<\infty:\qquad \qquad & \Xi(x,y;\kappa)=-\Xi(x,-y;\kappa)=-\Xi(x,y+\kappa;\kappa)=-\Xi(x,y+\kappa;\kappa).
\end{align*}
We also have monotonicity, $\Xi(x,y;\kappa)$ is non-increasing in $x$ for $y\in (0,\kappa)$ ($(0,\infty)$ when $\kappa=\infty$). For $\kappa=\infty$, we also have uniform  vertical limits,
\[
\lim_{y\to-\infty}\left(\Xi(x,y;\infty)-\theta(x;\infty)\right)=0,
\]
where $\theta$ is the solution from Proposition \ref{main_theorem:c_zero:1_leadsto_0} in the case $\kappa=\infty$.

%
\end{Theorem}
The proof of this result will be carried out in Sections \ref{section:H_infty} and \ref{section:H_problem}. 

\begin{Remark}[$\mathcal{O}$]\label{r:onon} There do not exist solutions asymptotic to oblique interfaces. More precisely, there do not exist solutions $u(x,y)$ with  
\[\lim_{x\to\infty}u(x,y)=0,\ \lim_{x\to -\infty} \left(u(x,y)-\bar{u}(\cos(\phi) x + \sin(\phi) y;L)\right)=0, \ \ u(x,y)=u(x,y+L),\ 0<\phi<\pi/2
.\]
This can be readily seen by noticing that the ``momentum'' $J$,  is constant in $x$, 
\[
J[u](x):=\dashint_y u_xu_y dy,\qquad \frac{d}{dx}J[u](x)=0
,\]
for solutions, hence $J$ is independent of $x$. Beyond a direct calculation, this can be seen by writing the elliptic equation as a first-order dynamical system, which then, due to the variational nature of the problem inherits a Hamiltonian structure. As a consequence, symmetries, in this case $y$-translations, are associated with conserved quantities. The momentum $J$ generates precisely the $y$-translations with respect to the standard symplectic structure $\omega((u_1,u_{1,x}),(u_2,u_{2,x})=\dashint u_1 u_{2,x}-u_2 u_{1,x} dx$. On the other hand, 
\[
J[u](+\infty)=0,\qquad J[u](-\infty)=\frac{\pi^2}{\kappa^2}\cos(\phi)\sin(\phi)\dashint \left(\bar{u}(\xi;\kappa)'\right)^2 d\xi\neq 0.
\] 

Similarly, there do not exist  solutions asymptotic to a single oblique interface, 
\[
\lim_{x\to -\infty} \left(u(x,y)-\tanh(\cos(\phi) x + \sin(\phi) y;\kappa)\right)=0,\ 0<\phi<\pi/2.
\]
We refer to \cite[\S 4.1]{lloyd} for another example where spatial symmetries and associated Hamiltonian conservation laws lead to explicit selection laws for patterns.
\end{Remark}

\begin{Remark}
The list of patterns mentioned here is connected through various limits. vertical stripes limit on $1\leadsto 0$ configurations, as the period $\kappa$ goes to $\infty$, in a locally uniform sense. Similarly, horizontal stripes $\mathcal{H}$-problem, limit on the  $\mathcal{H}_\infty$ solution, locally uniformly, or, when shifted in $y$ by $\kappa/2$, on the $1\leadsto 0$ solution, for $\kappa \to \infty.$ Similarly, oblique stripes limit on horizontal or vertical stripes, as the angle limits on $0$ and $\pi/2$.
\end{Remark}

\subsection{Main results: small speeds}\label{section:intro:c_greater_zero} We now state results concerning the existence and nonexistence of solutions to Allen-Cahn \eqref{Allen-Cahn} and Cahn-Hilliard \eqref{Cahn-Hilliard} with $c\gtrsim 0$. 

%

We start with the traveling-wave solutions to Allen-Cahn,
 \begin{eqnarray}\label{Allen-Cahn_c_greater_than_zero}
 -c u_x  = \Delta u + \mu(x) u  - u^3.
\end{eqnarray}

\begin{Proposition}[AC existence, $1\leadsto 0$, $\mathcal{H}$, $\mathcal{H}_\infty$; $c>0$]\label{t:acc}
The solutions found for $c=0$ can be continued smoothly to $c>0$. More precisely, there exist families of solutions to \eqref{Allen-Cahn_c_greater_than_zero} $\theta(x;c)$ for $0<c<\delta_1$ and $\Xi(x,y;\kappa,c)$, $\pi<\kappa\leq 
\infty$,  for $0\leq c<\delta_2(\kappa)$, with $\theta(x;0)=\theta(x)$ , $\Xi(x,y;\kappa,0)=\Xi(x,y;\kappa)$, where $\theta$ and $\Xi$ were found in Proposition \ref{main_theorem:c_zero:1_leadsto_0} and Theorem \ref{main_theorem:c_zero:H_infty}, respectively, satisfying the same limiting conditions as the solutions at $c=0$ for $x\to\pm\infty$. In the case $\mathcal{H}_\infty$, 
\[
\lim_{y\to\infty} \left(\Xi(x,y;\infty,c)-\theta(x;c)\right)=0.
\]
Moreover, the solutions depend smoothly on $c$, uniformly in $(x,y)$.
\end{Proposition}
The proof will be carried out in Section \ref{section:c_greater_zero:Allen_Cahn} and is based on the Implicit Function Theorem. 

We conclude the discussion of AC-dynamics with remarks on non-existence.
\begin{Remark}[AC non-existence, $\mathcal{O}$, $c>0$]\label{r:noo}
Oblique stripes are traveling waves in a coordinate system moving with speed $c_y$ in the $y$-direction, when $c\cos(\phi)+c_y\sin(\phi)=0$, so that they are solutions to 
\[ -c_y u_y -c u_x  = \Delta u + \mu(x) u  - u^3.
\]
The momentum $J$ introduced in Remark \ref{r:onon}, evaluated on a solution now solves 
\[
J_x=-cJ - c \dashint u_y^2,
\]
with limits $J(+\infty)=0$, $J(-\infty)=k_xk_y\dashint (\bar{u}')^2$. Solving the differential equation for $J$ with the boundary condition gives 
\[
J(x)=e^{-c(x-x_0)}J(x_0)+\int_{x_0}^x e^{-c(x-\xi)}\left(-c\dashint u_y^2(\xi,y)dy\right).
\]
Letting $x_0\to -\infty$ and using that $J$ is bounded gives
\[
J(x)=\int_{-\infty}^x e^{-c(x-\xi)}\left(-c\dashint u_y^2(\xi,y)dy\right),
\]
which, passing to the limit $x=-\infty$ or $x=+\infty$ gives a contradiction. 
\end{Remark}
Solutions to the $\mathcal{V}$-problem would naturally continue as periodic solutions to \eqref{Allen-Cahn} with temporal  period $T=2\pi/\omega$, $\omega=ck$, $k=\pi/\kappa$. 
\begin{Remark}[AC non-existence, $\mathcal{V}$; $c> 0$]\label{p:nodal}
There do not exist time-periodic solutions to \eqref{Allen-Cahn} with $c>0$ with spatial asymptotics
\[
\lim_{x\to -\infty} \left(u(t,x)-\bar{u}(x+ct;\kappa)\right)=0.
\]
To see this, we mimic the oblique case, Remark \ref{r:noo}. Define 
\[
J[u]=\dashint_t u_xu_t,
\]
for a time-periodic solution $u$. Then a short calculation gives 
\[
J_x=-cJ - c \dashint u_t^2,
\]
which leads to a contradiction in the same fashion as in Remark \ref{r:noo}.
\end{Remark}

We now turn to traveling-wave solutions of theCahn-Hilliard equation,
 \begin{eqnarray}\label{Cahn-Hilliard_c_greater_than_zero}
 - c u_x  = -\Delta\left(\Delta u + \mu(x) u  - u^3\right),
\end{eqnarray}
\begin{Proposition}[CH existence, $\mathcal{H}$; $c>0$]\label{t:chc}
The solutions found for $c=0$ can be continued smoothly to $c>0$. More precisely, there exist families of solutions to \eqref{Cahn-Hilliard_c_greater_than_zero}  $\Xi(x,y;\kappa,c)$, $\kappa\in (\pi,\infty)$,  for $0\leq c<\delta_2(\kappa)$, with $\Xi(x,y;\kappa,0)=\Xi(x,y;\kappa)$, where $\Xi$ was found in  Theorem \ref{main_theorem:c_zero:H_infty}, satisfying the same limiting conditions as the solutions at $c=0$ for $x\to\pm\infty$. Moreover, the solutions depend smoothly on $c$, uniformly in $(x,y)$.
\end{Proposition}

Solutions $1\leadsto 0$ and $\mathcal{H}_\infty$ cannot exist for Cahn-Hilliard due to mass conservation. 
\begin{Remark}[CH non-existence, $1\leadsto 0$, $\mathcal{H}_\infty$; $c>0$]\label{r:noch}
Mass conservation is exploited by integrating \eqref{Cahn-Hilliard_c_greater_than_zero},
\[
- c u  = -\left(u_{xx} + \mu(x) u  - u^3\right)_{x},
\]
using that $u\to 0$ for $x\to\infty$. As a consequence, $u=1$ is not an equilibrium of this ODE, such that $u\to 1$ for $x\to -\infty$ is not possible, hence existence of solutions with $1\leadsto 0$ is excluded for $c>0$. Since such solutions constitute boundary values at $\pm\infty$ for the $\mathcal{H}_\infty$-problem, we also conclude non-existence of solutions generating single interfaces. 
\end{Remark}

\paragraph{Outline.} 

We prove the results for $c=0$ in Sections \ref{section:1_leadsto_0}--\ref{section:H_problem}, starting with the $1\leadsto 0$ case as a technical warmup, followed by the $\mathcal{H}_\infty$ case, and the $\mathcal{H}$ case. The technical arguments here follow ideas from \cite{kolli2003approximation}. 
We address perturbation arguments that establish Propositions \ref{t:acc} and \ref{t:chc}, in Section \ref{section:c_greater_zero:Allen_Cahn}. We conclude in Section \ref{open_problems} with a brief discussion.

\paragraph{Notation.} In this paper we write $\mathbb{R}^+ := \{x \in \mathbb{R}| x>0\}$. $\mathscr{C}^k(X;Y)$, $\mathscr{C}_0^k(X;Y)$ and $\mathscr{C}^{(k,\alpha)}(X;Y)$ denote respectively, the space of $k$-times continuously differentiable functions, the space of $k$ times continuously differentiable functions with compact support in $X$, the space of $(k, \alpha)$ Holder continuously differentiable functions from $X$ to $Y$. We denote the Sobolev spaces over an open set $\Omega$  by $H^k(\Omega)$. The inner product of elements in a Hilbert space $\mathcal{H}$ is written as $\langle, \rangle_{\mathcal{H}}$. Norms on a Banach space $\mathcal{B}$ are denoted as $||\cdot||_{\mathcal{B}}$. The domain of an unbounded operator $\mathscr{L}$ is written $\mathcal{D}\left( \mathscr{L} \right)$. For a given operator $\mathscr{L}: \mathcal{D}(\mathscr{L}) \subset X \to Y$ we write $\mathrm{Ker}\left( \mathscr{L} \right) :=\{ u \in \mathscr{D}(\mathscr{L})| \mathscr{L}u=0 \}$ and $\mathrm{Rg}\left( 
\mathscr{L} \right) :=\{ f \in Y |\exists u \in \mathscr{D}(\mathscr{L}), Lu =f \}.$ A distribution $T \in \mathcal{
D}'(\Omega)$ satisfies  $T\geq 0$ in the sense of distributions if $T(\phi) \geq 0$ for any $\phi(\cdot) \in \mathcal{C}_K^{\infty}(\Omega; [0, \infty)).$ Let $T$ be a closed operator between Banach spaces $X$ and $Y$.

\paragraph{Acknoledgments.} R.M and A.S. are  grateful to the University of M\"unster, Germany, where part of this work carried out. R.M. also would like to thanks Itsv\'an Lagzi and Zolt\'an R\'acz from \"Eotvos University, Hungary, for stimulating discussions. R.M. acknowledges financial support through a DAAD Research Grant. A.S. acknowledges partial support throug  NSF grants DMS-1612441 and DMS-1311740.
\section{One-dimensional quenched patterns at zero speed and proof of Proposition \ref{main_theorem:c_zero:1_leadsto_0}}\label{section:1_leadsto_0}
We prove Proposition \ref{main_theorem:c_zero:1_leadsto_0}. We first outline a simple geometric proof, Section \ref{s:pp} based on phase plane analysis. We then, proceed with an alternative proof, based on comparison comparison principles and taking limits of domains of finite size in Sections \ref{s:21}--\ref{s:24}. This latter proof will carry over to the construction of solutions in the $\mathcal{H}$ and the $\mathcal{H}_\infty$ setting. 

\subsection{Phase plane analysis}\label{s:pp}

In order to prove Proposition \ref{main_theorem:c_zero:1_leadsto_0}, we study bounded solutions to the ODE
\begin{eqnarray}\label{allen_cahn_1d-eq1}
 \partial_x^2 u  + \mu(x) u - u^3 =0.
\end{eqnarray}
For $x>0$ and $x<0$, separately, solutions can readily be found by inspecting the phase portrait, which in turn is explicitly determined by the conserved Hamiltonians
\[
H^-(u,u_x)(x) = \frac{1}{2}u_x^2 +\frac{1}{2}u_2 -\frac{1}{4}u^4, \ x<0, \qquad 
H^+(u,u_x)(x) = \frac{1}{2}u_x^2 -\frac{1}{2}u_2 -\frac{1}{4}u^4, \ x>0.
\]
\begin{figure}[htb]
  \centering
    \includegraphics[width=0.9\textwidth]{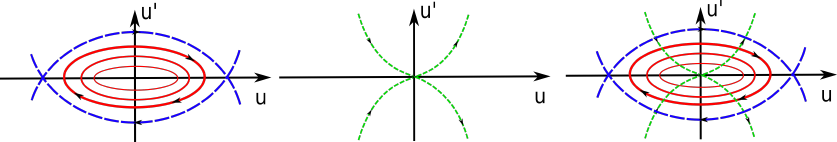}
  \caption{Phase portrait from level sets of $H^-$ (left, red and blue), $H^+$ (right, green) and  an overlay exhibiting intersections of the stable manifold and periodic orbits. .\label{phase_portraits_1D}}
\end{figure}
Continuity of $u$ and $u_x$ at $x=0$ implies that we find bounded solutions by intersecting level sets of $H^-$ and $H^+$ and solving from the intersection point backwards and forwards, respectively. Bounded solutions for $x>0$ lie in the level set $H^+=0$ and are given by the stable manifold of the origin. Figure \ref{phase_portraits_1D} shows that the stable manifold of the origin intersects the family of periodic orbits as well as the stable manifold of $\pm 1$, yielding the family of solutions described in Proposition \ref{main_theorem:c_zero:1_leadsto_0}. 

\subsection{The truncated problem} \label{s:21}

Of course, the phase plane arguments we relied on in the previous section do not obviously carry over to a two-dimensional PDE setup. We therefore pursue here an alternative strategy using PDE a priori estimates and comparison principles. We solve 
\begin{eqnarray}\label{intro_eq_1D_problem}
 u''  + \mu(x) u - u^3 =0,\ -M<x<L,\qquad u(-M)=1,\ u(L)=0,
\end{eqnarray}
for $0<M,L<\infty$, with continuity of $u$ and $u_x$ at $x=0$ as explained before. The idea then is to let $M\to\infty$ subsequently $L\to\infty$, to find that in the solution converges to a solution converging to $1$ and $0$ as $x\to\pm\infty$, respectively, thus proving the assertion in Proposition \ref{main_theorem:c_zero:1_leadsto_0} fvor $\kappa=\infty$; see the diagram below for a schematic representation. 
\begin{center}
\begin{tikzpicture}[>=stealth,every node/.style={shape=rectangle,draw,rounded corners},align=center,node distance=1cm]
    \node (full) {\begin{minipage}{3cm}\begin{center}
                   $ 1 \leadsto 0$-problem\\ in $\mathbb{R}$\\ \textbf{(solution $\theta(\cdot)$)}
                  \end{center} \end{minipage}};
    \node (ml) [below left=of full]{\begin{minipage}{4cm}\begin{center}
                                     Truncated \\$ 1 \leadsto 0$-problem \\in $(-M,L)$\\ \textbf{(solution $\theta_{(-M,L)}(\cdot)$)}
                                    \end{center} \end{minipage}};
    \node (infl) [below right=of full]{\begin{minipage}{4cm}\begin{center}
                                     Truncated \\$ 1 \leadsto 0$-problem \\in $(-\infty,L)$\\ \textbf{(solution $\theta_{(-\infty,L)}(\cdot)$)}
                                    \end{center} \end{minipage}};
    \draw[->] (full) to[out=180,in=45] node[draw=none,left,midway] {truncation} (ml);
    \draw[->] (ml) to[out=0,in=180] node[draw=none,above,midway] {$M \to \infty$} (infl)  ;
    \draw[->] (infl)to[out=135,in=0] node[draw=none,right,midway] {$L \to \infty$} (full);    
\end{tikzpicture}
\end{center}
The basic ideas of our approach are contained in the work of Kolli-Schatzmann \cite{kolli2003approximation}, where problems on a quarter plane were considered, with added difficulties due to the additional limit $L\to\infty$. Roughly following \cite{kolli2003approximation}, we define the following iterative scheme,
\begin{eqnarray}\label{H_1D_existence_sequence}
\left\{\begin{array}{c}
  -  \theta_{n+1}(x)'' + 5 \theta_{n+1}(x)  =  (5 +\mu(x)) \theta_n(x) - \theta_n^3(x), \quad \mbox{in} \, (-M,L) \\
 \theta_{n+1}(-M)=1, \quad \theta_{n+1}(L)=0, 
\end{array}\right.
\end{eqnarray}
We will start this iterative scheme with initial data $\theta_0(\cdot)$ from the class 
\begin{align*}
 \Psi_{1 \leadsto 0} &:= \left\{\theta \in \mathscr{C}^{(1,\alpha)}([-M,L])\,\middle|\, \theta(-M)=1,\ \theta(L)\geq 0,\ \theta(x)\in[0,1] \mbox{ for all } x \right\}\\
 & \quad \cap \left\{  \theta''(x) + \mu(x)\theta(x) - \theta^3(x) \leq 0, \   \theta''(x) + \mu(x)\theta(x) - \theta^3(x) \not\equiv 0, \, \mbox{in the sense of distributions}\right\}.
\end{align*}
Note $\Psi_{1\leadsto 0}\neq \emptyset$ since $\theta_0 \equiv 1 \in \Psi_{1 \leadsto 0}$. Moreover, the condition of $\theta_0(\cdot)$ not being a solution implies that $\theta_0(L) >0$ or that the distributional inequality $\theta_0''(x) + \mu(x)\theta_0(x) - \theta_0^3(x) \leq 0$ is not an equality.

Notice that the right hand side of \eqref{H_1D_existence_sequence} is non-decreasing as a function of $\theta_n$ for $\theta_n(\cdot) \in [0,1].$ Exploiting  uniform ellipticity of the operator $(-\partial_{xx} + 5)[\cdot]$ \cite[\S 8 \& 9]{brezis}, we readily find that 
the scheme represented  \eqref{H_1D_existence_sequence} is well defined and generates a unique sequence $\left(\theta_n(\cdot) \right)_{n \in \mathbb{N}}$ for every fixed initialization $\theta_0 \in \Psi_{1 \leadsto 0}$  .
In what follows we investigate some of the properties of this sequence.

\begin{Lemma}\label{Thm:properties_sequence_1d_scheme} For every $\theta_0 \in \Psi_{1\leadsto 0}$, the sequence $\left(\theta_j(\cdot) \right)_{j \in \mathbb{N}}$ of iterates from  \eqref{H_1D_existence_sequence} satisfies for all $j\geq 1$ that
\begin{enumerate}[label=(\roman*), ref=\theTheorem(\roman*)]
\hangindent\leftmargin
 \item \label{Thm:properties_sequence_1d_scheme:prop_a}  $0 < \theta_{(-M,L)}(x) <\theta_{j+1}(x)<\theta_j(x) <1$ for all $x \in (-M, L)$ and any solution $\theta_{(-M,L)}$of \eqref{intro_eq_1D_problem};
 %
 \item \label{Thm:properties_sequence_1d_scheme:prop_d}the sequence $\left(\theta_j(\cdot) \right)_{j\geq 0}$ is precompact in $\mathscr{C}^{1,\alpha}(-M,L),$ $\mbox{ for all } 0\leq\alpha <1$.
\end{enumerate}
\end{Lemma}
\begin{Proof}
All  results follow by induction and successive applications of the maximum principle; see for instance \cite[\S 3]{gilbarg2015elliptic}. Assume without loss of generality that $\theta_0(L) >0$. We begin by  proving (i) for $j=1$. 
We use \eqref{H_1D_existence_sequence} and the fact  $\theta_0\in \Psi_{1 \leadsto 0}$ to conclude that  $ -(\theta_0 - \theta_1)'' + 5(\theta_0 - \theta_1) \geq 0$ in the sense of distributions. We know that $\theta_0 - \theta_1 \geq 0$ on $\partial(-M,L)$, and that $\theta_0 \not\equiv \theta_1$ since $\theta_0$ is not a solution. We shall prove that $\theta_0 >\theta_1$ in $(-M,L)$. We first show that $\theta_0 \geq \theta_1$ in $(-M,L)$: by construction we have  that $\theta_0 - \theta_1 \in \mathcal{C}^{(1,\alpha)}([-M,L];[0,1]) \subset H^1((-M,L))$. Using   ellipticity of $(- \partial_x^5 +5)[\cdot]$ and the maximum principle \cite[\S 8, Thm. 8.1]{gilbarg2015elliptic} we conclude that   $\sup_{(-M,L)}[-(\theta_0 - \theta_1)]\leq \sup_{\partial(-M,L)}[-(\theta_0 - \theta_1)]$ as claimed. We now argue by contradiction to show that $\theta_0 > \theta_1$ in $(-M,L)$.  Assume that there exists a $x_0 \in (-M,L)$ such that $\left(\theta_0 - \theta_1 \right)(x_
0)=0$ and let $ B \subset \subset (-M,L)$ be an open ball centered at $x_0$.  Again by the maximum principle \cite[\S 8, Thm. 8.19]{gilbarg2015elliptic} applied  to  $-(\theta_0 - \theta_1)$ we conclude that
\[
\sup_{ B}\left(-(\theta_0 - \theta_1)\right) = \sup_{\partial(-M,L)}\left(-(\theta_0 - \theta_1)\right) ,
\]
such that $(\theta_0 - \theta_1)$ is a constant in $(-M,L)$, which however is incompatible with continuity of $(\theta_0 - \theta_1)(\cdot)$ and its boundary conditions. By induction, we conclude pointwise monotonicity and $\theta_j(x)\in(0,1)$. Following the same reasoning, one readily concludes strict positivity of any solution $\theta_{(-M,L)}$, which concludes the proof of (i).

To show (ii), notice that the sequence $\theta_j$ bounded in $W^{2,p}$ for any $p<\infty$. Using the compactness of the embedding into $
\mathscr{C}^{1,\alpha}$ gives the desired precompactness. 

\end{Proof}

The next result gives the uniqueness of the solution to problem \ref{intro_eq_1D_problem}. 
\begin{Lemma}[Uniqueness to truncated $1 \leadsto 0$] There exists at most one solution to problem \eqref{intro_eq_1D_problem}.\end{Lemma}
\begin{Proof}
The proof here is similar to \cite[Lem. 1.2]{kolli2003approximation}. Assume there are  two solutions, $\theta(\cdot)$, $\tilde{\theta}(\cdot)$ so that $\theta (\cdot)\not\equiv \tilde{\theta}(\cdot)$. Define the set  $\mathscr{D} = \{x \in [-M,L] \, |\,  \theta(x) \neq \tilde{\theta}(x) \}.$  Now $\mathscr{D}$ is open by continuity of $\theta$ and we choose a connected component $(a,b)$, such that $\theta(x)> \tilde{\theta}(x)$, $x \in (a,b)$, $\theta(x)> \tilde{\theta}(x)$, $x\in\{a,b\}$. Now, since both $\theta$ and $\tilde{\theta}$ are solutions, we can integrate against test functions $\theta$ and $\tilde{\theta}$ on the interval $(a,b)$\footnote{Notice that this is not direct, since these function don't solve the PDE in the classical sense and  distributions are applied to the space of smooth compactly supported functions. However, we know that distributions with finite order (say, order $k$) can be extended to the space of $\mathscr{C}_0^k$ functions (cf. \cite[\S 2]{Hormander}).}, which gives
 \begin{align*}
\int_a^b  (- \theta'' \tilde{\theta} + \tilde{\theta}'' \theta )dx+ \int_a^b  (\theta^2 -  \tilde{\theta}^2) \theta \tilde{\theta} dx &=0  \implies (- \theta' \tilde{\theta} + \tilde{\theta}' \theta )\big|_a^b+ \int_a^b  (\theta^2 -  \tilde{\theta}^2) \theta \tilde{\theta} dx =0, 
 \end{align*}
where the first term was integrated by parts. Since $\theta > \tilde{\theta}$ in $(a,b)$, $\theta(x) = \tilde{\theta}(x)$ for $x \in \{a,b\}$  and we find that the first term is non-negative. The second term is however  strictly positive, thanks to Proposition \ref{Thm:properties_sequence_1d_scheme}(i) and the assumption that $\theta > \tilde{\theta}$ in $(a,b)$, thus proving the result by contradiction. 
\end{Proof}

\begin{Proposition}[Existence to truncated $1 \leadsto 0$-problem]\label{existence_lemma:1_leadsto_0} There exists a unique solution  $\theta_{(-M,L)}(x)$ to problem \eqref{intro_eq_1D_problem}.
\end{Proposition}
\begin{Proof}
Since $\left(\theta_n(\cdot) \right)_{n \in \mathbb{N}}$ is monotone, we can define $\theta_{(-M,L)} := \displaystyle{\inf_{n\in \mathbb{N}}} \left(\theta_n(\cdot) \right)$ which is clearly measurable. As each term in this sequence  is bounded uniformly by $1$, pointwise convergence implies convergence in $L^1$, which in turn implies that the limiting function is a solution in the sense of distributions, satisfying the boundary conditions due to pointwise convergence. 
\end{Proof}
Note that the infimum of an iteration \eqref{H_2D_existence_sequence_H_infty} is independent of the choice of the starting point in $\Psi_{1 \leadsto 0}$. Slightly stronger then the simple uniqueness result, this is the basis of the comparison methods we shall employ later, when classical strong maximum principles do not apply directly due to lack of sufficient regularity.
\subsection{Properties of solutions to the truncated problem}\label{s:23}
In this section we study the qualitative properties of $\theta_{(-M,L)}(x)$ as we vary $x$, $M$ and $L$ separately. We start with a lemma, which we appeal to  repeatedly.

\begin{Lemma}[Comparison principles, $1 \leadsto 0$-problem]\label{Thm:sub-supersolutions_1_leadsto_0}Let $\theta_{(-M,L)}(\cdot,\cdot)$ be the solution from Proposition \ref{existence_lemma:1_leadsto_0}.
 
\begin{enumerate}[label=(\roman*), ref=\theTheorem(\roman*)]
\hangindent\leftmargin
\item \emph{(1D supersolutions)}\label{Thm:sub-supersolutions_1_leadsto_0:lemma_a} Suppose that $v$ satisfies, in the sense of distributions,
\begin{eqnarray}\label{Thm:sub-supersolutions_1_leadsto_0:lemma_a:inequality_1}
v''(x) + \mu(x)v(x) - v^3(x) \leq 0,\quad  0 \leq v\leq 1,  \quad v(-M) =1, \quad  v(L) \geq 0. 
\end{eqnarray}
Then $v(x) \geq \theta_{(-M,L)}(x)$ for $x \in (-M, L).$ In particular, $v(x) \geq \theta_{(-M,L)}(x)$ for any solution $v$ of 
\begin{eqnarray}\label{Thm:sub-supersolutions_1_leadsto_0:lemma_a:inequality_2}
v''(x) + v(x) - v^3(x) = 0,\quad  0 \leq v\leq 1,  \quad v(-M) =1, \quad v(L) \geq 0. 
\end{eqnarray}

\item \emph{(1D subsolutions)}\label{Thm:sub-supersolutions_1_leadsto_0:subsolution_lemma:1_leadsto_0} Suppose that $v$ satisfies, in the sense of distributions,
\begin{eqnarray*}
w''(x) + \mu(x)w(x) - w^3(x) \geq 0,\quad  0\leq w\leq 1,  \quad w(-M) \leq 1, \quad w(L) = 0.  
\end{eqnarray*}
Then $w(x) \leq \theta_{(-M,L)}(x)$ for $x \in (-M, L).$ In particular, $w(x) \leq \theta_{(-M,L)}(x)$ for any solution $w$ of 
\begin{eqnarray*}
w''(x) - w(x) - w^3(x) = 0,\quad  0 \leq w\leq 1, \quad w(-M) \leq1, \quad w(L) = 0. 
\end{eqnarray*}
\end{enumerate}
 
\end{Lemma}

\begin{Proof}
Inequality \eqref{Thm:sub-supersolutions_1_leadsto_0:lemma_a:inequality_2} is an easy consequence of inequality \eqref{Thm:sub-supersolutions_1_leadsto_0:lemma_a:inequality_1}: indeed, if $v$ satisfies \eqref{Thm:sub-supersolutions_1_leadsto_0:lemma_a:inequality_2} then
$$v''(x) + v(x) - v^3(x) = 0 \Longleftrightarrow  v''(x) + \mu(x)v(x) - v^3(x) = (\mu(x) -1) v(x) \leq 0.$$
To prove (i), we assume  that $v \in \Psi_{1 \leadsto 0}$ since otherwise  $v = \theta_{(-M,L)}$ by uniqueness. Setting $\theta_0 \equiv v$ and using Lemma \ref{Thm:properties_sequence_1d_scheme}(i)  now implies the result. 
 
To prove (ii), set $\theta_0 \equiv 1$. Notice that $\theta_0 \in \Psi_{1 \leadsto 0}$ and that $w \leq \theta_0$ by definition of $w$.  By induction, following the proof of Lemma \ref{Thm:properties_sequence_1d_scheme}(i), it follows that $\left(\theta_j\right)_{j \in \mathbb{N}}$ satisfies $\theta_j \geq w$, hence $\displaystyle{\inf_{j \in \mathbb{N}}} \theta_j = \theta_{(-M,L)} \geq w.$
\end{Proof}
In order to compare the families of solutions as $M,L$ vary, we construct trivial constant extensions of functions $u$ defined on $(-M,L)\to \R$, formalized as an operator $\mathscr{E}$, 
\begin{eqnarray*}
 \mathscr{E}\left[u\right](x) = \left\{ \begin{array}{ccc}
                        u(x), & \quad & \mbox{for} \quad x \in (-M, L)\\
                        1, &  \quad & \mbox{for}\quad  x \leq -M\\
                         0, &  \quad & \mbox{for} \quad  x \geq L.
                       \end{array} \right.
\end{eqnarray*}
\begin{Lemma}[Properties of the extension]\label{Thm:monotonicity_properties_1_to_0_1_to_0} The following properties of $\mathscr{E}[\theta_{(-M,L)}](\cdot)$ hold.
\begin{enumerate}[label=(\roman*), ref=\theTheorem(\roman*)]
\hangindent\leftmargin
 \item\emph{(Monotonicity of $\mathscr{E}$)}\label{Thm:monotonicity_properties_1_to_0:monotonicity_extension_operator}	
 We have $0 \leq \mathscr{E}\left[\theta_{(-M,L)}\right](\cdot) \leq 1$. Furthermore, for $w$ defined on a subset $A$, $ (-M,L) \subset A \subset (-\infty,L)$ with  $0 \leq w(\cdot) \leq 1$ and $ 0 \leq w(\cdot) \leq \theta_{(-M,L)}(\cdot)$ in $(-M,L)$, we have
 %
$0 \leq \mathscr{E}\left[w\right](\cdot) \leq \mathscr{E}\left[\theta_{(-M,L)}\right](\cdot)  \, \, \mbox{on} \, \, \mathbb{R}.   $                                                                                                                                                                                                                          
 
 \item \emph{(Monotonicity in $M$)}\label{Thm:monotonicity_properties_1_to_0:prop_monotonicity_in_M}
Let  $0 \leq M <\widetilde{M}$ and $L \geq 0$ be fixed. Then $\mathscr{E}\left[\theta_{(-\widetilde{M},L)}\right](x) \leq  \mathscr{E}\left[\theta_{(-M,L)}\right](x).$

\item \emph{(Monotonicity in $L$)}\label{Thm:monotonicity_properties_1_to_0:prop_monotonicity_in_L}
Let  $0 \leq L <\widetilde{L}$ and $M \geq 0$ be fixed. Then $\mathscr{E}\left[\theta_{(-M,L)}\right](x) \leq  \mathscr{E}\left[\theta_{(-M,\widetilde{L})}\right](x).$
\item \emph{(Monotonicity in $x$)}\label{Thm:monotonicity_properties_1_to_0:1D_monotonicity_in_x} For every fixed $M$ and $L$ the mapping $x\mapsto \mathscr{E}\left[\theta_{(-M,L)}\right](x) $ is non-increasing.
\end{enumerate}
\end{Lemma}

\begin{Proof}
The first assertion (i) is immediate from the definition of $\mathscr{E}$.  
To show (ii), we use  Lemma \ref{Thm:properties_sequence_1d_scheme}(i) together with  Lemma \ref{Thm:sub-supersolutions_1_leadsto_0}(ii) to conclude that 
$$
v(x) :=\mathscr{E}\left[\theta_{(\widetilde{M},L)}\right]\big|_{[-M,L]}(x) = \theta_{(\widetilde{M},L)}\big|_{[-M,L]}(x)
$$ 
is a subsolution to \eqref{intro_eq_1D_problem}. Hence, using again Lemma \ref{Thm:sub-supersolutions_1_leadsto_0}(ii),  $ v(x) \leq \theta_{(-M, L)}(x)$, for $x  \in (-M,L),$ and, applying (i), we obtain the result. The proof of (iii) is analogous.

To prove (iv), fix $x_0 >0$ and pick $L >x_0$, $M > x_0$. Notice that  $w(x) := \theta_{(-M,L)}(x+x_0)$ is well defined in $x\in (-M, L-x_0).$ We claim that $w(\cdot)$ is a subsolution to the problem \eqref{intro_eq_1D_problem} in $(-M, L-x_0).$ Indeed, on the boundary $x= -M$ we have
 $$ w(-M) = \theta_{(-M,L)}(-M+x_0) \leq 1 = \theta_{(-M,L-L_0)}(-M),$$
and $$ w(L-x_0) = \theta_{(-M,L)}(L-x_0 +x_0) = \theta_{(-M,L)}(L) = 0 =  \theta_{(-M,L-L_0)}(L - x_0).$$
It remains to show that $w(\cdot)$ satisfies  Lemma \ref{Thm:sub-supersolutions_1_leadsto_0}(ii) in $(-M,L-x_0)$. We have that
\begin{align*}
  w''(x) + \mu(x) w(x) - w(x)^3 &=\left\{ w''(x) + \mu(x+x_0) w(x) - w^3(x) \right\} \\
  &\quad +\left(\mu(x) - \mu(x+x_0)\right)w(x+x_0) =:I_1 + I_2.
\end{align*}
 First note that $I_1 = 0$ in the sense of distributions, by definition of $w(\cdot)$. Next, $I_2 \geq 0$ in the sense of distributions, since $\mu(x) \geq \mu(x + x_0)$, for $x_0 >0$, and $w(\cdot)$ is non-negative. We conclude that $w(x)  \leq \theta_{(-M,L-x_0)}(x),$ due to Lemma \ref{Thm:sub-supersolutions_1_leadsto_0}(ii). Now the result follows, since by  (iii), 
\begin{eqnarray*}
 w(x) := \theta_{(-M,L)}(x+x_0) \leq \theta_{(-M,L - x_0)}(x){\leq} \theta_{(-M,L)}(x).  
 \end{eqnarray*}
\end{Proof}

The next result is relevant only for the solutions to the  $\mathcal{H}$  and $\mathcal{H}_{\infty}$-problems.  
\begin{Lemma}[Continuous dependence of $\theta_{(-M,L)}(\cdot)$ on $L,M$]\label{1D_lemma_continuity_in_L_M}
Let $0<M< \infty$, $0 <L< \infty$. The mappings $L\mapsto \mathscr{E}\left[ \theta_{(-M,L)}\right ](\cdot)$ and $M\mapsto \mathscr{E}\left[ \theta_{(-M,L)}\right ](\cdot)$ are continuous in the sup norm.
\end{Lemma}
\begin{Proof} 
Fix  $\epsilon >0$. First we prove continuity from the right in $L$. Since the extension operator extends the functions $\theta_{(-M,L)}(\cdot)$ as uniformly continuous functions in $\mathbb{R}$ the result will follow if we show that 
\begin{eqnarray}\label{continuity_inequality_in_L}
 \mathscr{E}\left[\theta_{(-M,L) }\right](x) \leq \mathscr{E}\left[\theta_{(-M,L+\epsilon) }\right](x) \leq  \mathscr{E}\left[\theta_{(-M+ \epsilon,L+\epsilon) }\right](x) \leq \mathscr{E}\left[\theta_{(-M,L) }\right](x - \epsilon) 
\end{eqnarray}
 for $x \in (-M,L +\epsilon)$. Indeed, \eqref{continuity_inequality_in_L} implies that
 \begin{equation*}
  |\mathscr{E}[\theta_{(-M,L+\epsilon)}(x)] - \mathscr{E}[\theta_{(-M,L)}](x)| \leq |\mathscr{E}[\theta_{(-M,L)}](x-\epsilon) - \mathscr{E}[\theta_{(-M,L)}](x)|
 \end{equation*}
and the result follows from uniform continuity of the function $\mathscr{E}[\theta_{(-M,L)}][\cdot]$ in $\mathbb{R}$.
\vspace{\lineskip}

 The first two inequalities in \eqref{continuity_inequality_in_L} are clearly true, due to monotonicity of $\theta_{(M,L)}(\cdot)$ in $L$ and $M$ thus we focus our attention on the last inequality.   It turns out that the latter is clearly true in $(-M, -M + \epsilon)$ so it suffices to show that the inequality
\begin{align*}
\mathscr{E}\left[\theta_{(-M+ \epsilon,L+\epsilon) }\right]\big|_{(-M+ \epsilon, L + \epsilon)}(x) &= \theta_{(-M+ \epsilon,L+\epsilon) }(x) \\
&\leq  \mathscr{E}\left[\theta_{(-M,L) }\right]\big|_{(-M+ \epsilon, L + \epsilon)}(x - \epsilon) = \theta_{(-M,L) }(x - \epsilon)
\end{align*}
holds true in $x \in (-M + \epsilon,L +\epsilon).$ We prove this by showing that $u(\cdot) := \theta_{(-M,L) }(\cdot - \epsilon)$ is a supersolution to problem \eqref{intro_eq_1D_problem} in the interval $(-M + \epsilon, L+ \epsilon).$ Indeed, $ u(-M + \epsilon) = 1$ and $u(L + \epsilon) = 0.$ Furthermore, 
\begin{align*}
u''(x) + \mu(x) u(x) - u^3(x) &=   \theta_{(-M,L) }''(x - \epsilon) + \mu(x) \theta_{(-M,L) }(x - \epsilon) - \theta_{(-M,L) }^3(x - \epsilon)\\
                 &\leq  \left\{ \theta_{(-M,L) }''(x - \epsilon) + \mu(x - \epsilon) \theta_{(-M,L) }(x - \epsilon) - \theta_{(-M,L) }^3(x - \epsilon) \right\}\\
                 & +\left\{[\mu(x ) -  \mu(x - \epsilon)] \theta_{(-M,L) }(x - \epsilon) \right\} = \mathscr{J}_1 + \mathscr{J}_2.
\end{align*}
First note that $\mathscr{J}_1 =0$ in the sense of distributions, since $\theta_{(-M,L) }(\cdot)$ is a solution to problem \eqref{intro_eq_1D_problem} in $(-M,L)$. Next, $\mathscr{J}_2 \leq 0$ because $x \mapsto \mu(x)$ is non-increasing. Now apply Lemma \ref{Thm:sub-supersolutions_1_leadsto_0}(i) to conclude the argument.
Analogously, one can show that 
\begin{eqnarray}\label{continuity_inequality_in_L_left}
 \mathscr{E}\left[\theta_{(-M,L) }\right](x +\epsilon) \leq \mathscr{E}\left[\theta_{(-M - \epsilon,L -\epsilon) }\right](x) \leq  \mathscr{E}\left[\theta_{(-M,L -\epsilon) }\right](x) \leq  \theta_{(-M,L) }(x),
\end{eqnarray}
in $x \in (-M-\epsilon,L-\epsilon)$, which proves the continuity in $L$ from the left and therefore continuity in $L$. The proof of the result for $M$ is analogous.
\end{Proof}
\begin{Corollary}\label{1D_continuity_in_L_M}
 The mapping $L\mapsto \mathscr{E}\left[ \theta_{(-\infty,L)}\right ](\cdot)$ is continuous in the sup norm on $0<L,\infty$.
\end{Corollary}
\begin{Proof}
Fix  $\epsilon >0$. Right continuity is a consequence of inequality \eqref{continuity_inequality_in_L} after taking the infimum in $M$; analogously, one can prove continuity on the left using \eqref{continuity_inequality_in_L_left}.
\end{Proof}

\subsection{Passing to the limit}\label{s:24}

We are now ready to pass to the limit $M=\infty$ and subsequently prove Proposition \ref{main_theorem:c_zero:1_leadsto_0} by letting $L\to\infty$. Define 
 \begin{eqnarray}\label{minuns_infty_L_1D}
\theta_{(-\infty,L)}(x) := \inf_{M >0}\mathscr{E}\left[\theta_{(-M,L)}\right](x)  = \lim_{M \to \infty}\mathscr{E}\left[\theta_{(-M,L)}\right](x),
 \end{eqnarray}
where the last equality is a consequence of Lemma \ref{Thm:monotonicity_properties_1_to_0_1_to_0}(i).
 \begin{Proposition}\label{Thm:properties_1_leadsto_0}
 The following properties hold for the family $\theta_{(-\infty,L)}$.
\noindent
 \begin{enumerate}[label=(\roman*), ref=\theTheorem(\roman*)]
 \item {(Monotonicity)}\label{1D_monotonicity_infty_L}
 The functions $x\mapsto \theta_{(-\infty,L)}(x)$ are defined for every $x \in \mathbb{R}.$
  The mapping  $L \mapsto \theta_{(-\infty,L)}(x)$ is non-decreasing for any fixed x. Furthermore, the mapping  $x \mapsto \theta_{(-\infty,L)}(x)$ is non-decreasing  for any fixed L.

 \item\label{minuns_infty_L_1D_theorem}The function $\theta_{(-\infty,L)}(x)$ solves   \eqref{intro_eq_1D_problem} on  $(-\infty, L).$ Furthermore, $\displaystyle{\lim_{x \to -\infty}}\theta_{(-\infty,L)}(x) = 1.$
   \end{enumerate}   
 \end{Proposition}
 \begin{Proof}
To prove (i), recall that the 
family of functions $\left\{ \mathscr{E}\left[\theta_{(-M,L)}\right](\cdot)\right\}_{M,L}$ is uniformly bounded, pointwise non-increasing in $M$ and non-decreasing in $L$, and decreasing in $x$ for fixed $(M,L)$. Passing to the limit $M=\infty$ by taking the infimum, we conclude that $\left\{ \mathscr{E}\left[\theta_{(-\infty,L)}\right](\cdot)\right\}_{L}$  is non-decreasing in $L$ and non-increasing in $x$. 

To prove (ii), notice that pointwise convergence and boundedness of the sequence $\left(\mathscr{E}(\theta_{(-M, L)})(\cdot)\right)_{M >0}$ implies convergence  in the sense of distributions to a weak solution, which clearly satisfies the boundary condition at $x=L$ by pointwise convergence. It remains to show that  It remains to show that
\begin{equation}\label{boundary_condition_at_minus_infty}
 \lim_{x \to -\infty}\theta_{(-\infty, L)}(x) = 1.
\end{equation} 
Define  $v(x):= \theta_{(-\infty,0)}(x)=-\tanh(x/\sqrt{2})$. Applying the classical strong maximum principle \cite[\S 3] {gilbarg2015elliptic} in the interval $(-M,0)$ we obtain $v\big|_{[-M,0]}(\cdot) \leq \theta_{(-M,L)}(\cdot) \quad \mbox{for} \quad  x\in (-M,0).$
Since $\theta_{(-M,L)}(\cdot)$ is non-negative, $\max\left\{v\big|_{[-M,L]}(\cdot), 0\right\}  \leq \theta_{(-M,L)}(\cdot)$ for $x \in (-M,L).$ Lemma \ref{Thm:monotonicity_properties_1_to_0_1_to_0}(i) implies that
$$ 
\max\left\{v(x), 0\right\}  \leq \mathscr{E}\left[\max \{v\big|_{[-M,0]}(\cdot), 0 \} \right](x)\leq \mathscr{E}\left[\theta_{(-M,L)}\right](x), \quad \mbox{ for all } M >0,
$$
and we conclude that $ \max\left\{v(x), 0\right\}  \leq \inf_{M \in \mathbb{N}} \mathscr{E}\left[\theta_{(-M,L)}\right](x) =: \theta_{(-\infty,L)}(x).$ Since clearly  $\displaystyle{\lim_{x \to -\infty}} v(x) = 1$ and $\theta_{(-\infty,L)}(x) \leq 1$, we obtain the limit \eqref{boundary_condition_at_minus_infty}. 
\end{Proof} 
 
We are now ready to prove our main result in one space dimension.
 
\begin{Proof}[of Proposition \ref{main_theorem:c_zero:1_leadsto_0}]
Let $\theta(x) := \sup_{L >0} \theta_{(-\infty,L)}(x) =  \displaystyle{\lim_{L \to \infty}} \theta_{(-\infty,L)}(x).$
We claim that the function $\theta(\cdot)$ solves the \textbf{$1\leadsto 0$} problem  in $\mathbb{R}$. Notice that $\displaystyle{\lim_{x \to -\infty}\theta} =1$  follows from Proposition \ref{Thm:properties_1_leadsto_0}(ii). The asymptotic behavior for $x \to +\infty$ follows from  comparison principles. Indeed, define the  function 
 $$w(x ) = \frac{1}{\sqrt{2}} \mathrm{csch}(x+x_0), \quad \mbox{for} \quad x>0,\quad  x_0 >0, $$
 such that $w(0) < 1$. One readily verifies that for $M$ sufficiently small, $w(\cdot)$ is a supersolution of \eqref{H_1D_existence_sequence} in $(-M, L)$. The smallness assumption on $M$ is used to assure that $0\leq w \leq 1$, while $L$ can be arbitrary. Since $w$ satisfies all the properties in Corollary  \eqref{Thm:sub-supersolutions_1_leadsto_0:lemma_a:inequality_2}, we obtain $\theta_{(-M,L)}(x) \leq w(x) $ for any $M, L$, $ x > 0.$  We take (in this order) the infimum  in $M$ and the supremum  in $L$  to obtain $\theta(x) \leq w(x)$ for $x >0,$ which establishes the desired asymptotics. Monotonicity in $x$ is preserved by taking the ordered supremum of monotone functions, which concludes the proof.  
\end{Proof}


%
\section{Two-dimensional quenched patterns at zero speed --- one interface}\label{section:H_infty}
In this section, we shall prove Theorem \ref{main_theorem:c_zero:H_infty} in the case $\kappa =\infty$. We first reduce the problem to a half plane and truncate in an analogous fashion to Section \ref{s:21} in Section \ref{section:symmetries}, reducing to a problem in a rectangle $\Omega_{(-M,L)} := (-M,L)\times(-M, 0)$. 
We then let the truncation size go two infinity in two steps, following again the strategy from Section \ref{section:1_leadsto_0}, first establishing monotonicity properties and letting $M\to\infty$ in Section \ref{s:32}, then establishing properties of solutions in $\Omega_{(-\infty,L)}$ and letting $L\to\infty$ in Sections   \ref{properties_theta_minusinfty_L}   and \ref{H_infty_problem_solution}. The diagram below illustrates the strategy in analogy with the one-dimensional case. 
\begin{center}
\begin{tikzpicture}[>=stealth,every node/.style={shape=rectangle,draw,rounded corners},align=center,node distance=1cm]
    \node (full) {\begin{minipage}{4.5cm}\begin{center}
                   $\mathcal{H}_{\infty}$-problem in\\ $\Omega_{(-\infty, \infty)} = \mathbb{R}\times(-\infty,0)$\\ \textbf{(solution $\Theta(\cdot)$)}
                  \end{center} \end{minipage}};
    \node (ml) [below left=of full]{\begin{minipage}{4cm}\begin{center}
                                     Truncated \\$\mathcal{H}_{\infty}$-problem\\in $\Omega_{(-M, L)}$\\ \textbf{(solution $\Theta_{(-M,L)}(\cdot)$)}
                                    \end{center} \end{minipage}};
    \node (infl) [below right=of full]{\begin{minipage}{4cm}\begin{center}
                                     Truncated \\$\mathcal{H}_{\infty}$-problem \\in $\Omega_{(-\infty, L)}$\\ \textbf{(solution $\theta_{(-\infty,L)}(\cdot)$)}
                                    \end{center} \end{minipage}};
    \draw[->] (full) to[out=180,in=45] node[draw=none,left,midway] {truncation} (ml);
    \draw[->] (ml) to[out=0,in=180] node[draw=none,above,midway] {$M \to \infty$} (infl)  ;
    \draw[->] (infl)to[out=135,in=0] node[draw=none,right,midway] {$L \to \infty$} (full);    
\end{tikzpicture}
\end{center}
\subsection{Reducing and truncating the domain}\label{section:symmetries}

In this chapter, we write $\Theta(x,y):=\Xi(x,y;\infty)$, suppressing thereby the extra parameter $L=\infty$. 
The problem for $\Theta$, posed  on $\R^2$ \eqref{ach0} reads 
\begin{equation}\label{e:hinf}
\Delta \Theta + \mu(x) \Theta - u^3,\quad \lim_{x\to -\infty}\Theta(x,y)=-\tanh(y/\sqrt{2}),\quad \lim_{y\to 
\pm\infty}\Theta(x,y)=\mp\theta(x),
\end{equation}
where $\theta$ is the one-dimensional pure phase selecting solution. 

We start by exploiting the fact that the nonlinearity is odd by solving for $\Theta$ on $\R\times (-\infty,0)$ with Dirichlet boundary conditions at $y=0$. Setting $\Theta(x,-y):=-\Theta(x,y)$ for $y>0$ then readily gives the desired solution on $\R^2$. As an extra benefit, this reduction removes the non-uniqueness of solutions induced by translation invariance. More importantly, it avoids a weak indefiniteness in the problem induced by the potential bending of the interface emerging in $x<0$. 

As a next step, we truncate the problem, setting up the \emph{truncated} $\mathcal{H}_\infty$-problem, 
\begin{align}
\Delta u + \mu(x) u - u^3&=0, & (x,y)\in\Omega_{(-M,L)},\nonumber\\
u&=g_{(-M,L)},& (x,y)\in\partial\Omega_{(-M,L)},\label{2D_H_infty_problem}
\end{align}
where $g_{(-M,L)}(x,y):=\theta_{(-M,L)}(x)\cdot \theta_{(-M,0)}(y)$,   $\theta_{(-M,L)}(\cdot)$ the solutions to the truncated one-dimensional problem  \eqref{intro_eq_1D_problem} on the interval $(-M, L)$. Here, as mentioned in the introduction, we refer to a solution in the sense of distributions, since solutions are not smooth across $x=0$. Exploiting that $0\leq u\leq 1$ and  Agmon-Douglis-Nirenberg regularity, we readily conclude that $u$ and derivatives are H\"older continuous across $x=0$, and, in fact, $u\in \mathscr{C}^{(1,\alpha)}(\overline{\Omega_{(-M,L)}}), $  $\mbox{ for all } 0\leq\alpha <1$.
%

We construct unique solutions to this truncated problem by iterating the following scheme,
\begin{eqnarray}\label{H_2D_existence_sequence_H_infty}%
\left\{\begin{array}{c}
  -  \Delta \Theta_{n+1}+ 5 \Theta_{n+1} =  (5 +\mu(x)) \Theta_n - \Theta_n^3  \\
 (\Theta_{n+1} - g_L)\big|_{\partial \Omega_{(-M,L)}}=0
\end{array}\right.
\end{eqnarray}
where $\Theta_0(\cdot)$ is chosen in the class 
\begin{align*}
 \Psi_{\mathcal{H}_{\infty}} &:= \left\{\Theta \in \mathscr{C}^{(1,\alpha)}(\Omega_{(-M,L)})\,\middle|\,(\Theta_{0} - g_{-M,L})\big|_{\partial \Omega_L} \geq 0,\ \Theta(x)\in[0,1] \mbox{ for all } x,y \right\}\\
 & \quad \cap \{  \Delta\Theta + \mu(x)\Theta- \Theta^3 \leq 0, \, \  \Delta\Theta + \mu(x)\Theta - \Theta^3 \not\equiv 0 \, \mbox{in the sense of distributions}\}.
\end{align*}
%

%
%
\begin{Lemma}\label{Thm:properties_sequence_H_infty} For every $\Theta_0 \in \Psi_{\mathcal{H}_{\infty}}$, the sequence of iterates $\left(\Theta_j(\cdot) \right)_{j \in \mathbb{N}}$ from \eqref{H_2D_existence_sequence_H_infty} satisfies for all $j\geq 1$that 
\begin{enumerate}[label=(\roman*), ref=\theTheorem(\roman*)]
  \item \label{Thm:properties_sequence_2d_scheme:prop_a}  $0 < \Theta_{(-M,L)}(x,y) <\Theta_{j+1}(x,y)<\Theta_j(x,y) <1$ for all $(x,y) \in (-M, L)$ and any solution $\Theta_{(-M,L)}$of \eqref{intro_eq_1D_problem};
 %
 \item \label{Thm:properties_sequence_2d_scheme:prop_d}the sequence $\left(\Theta_j(\cdot) \right)_{j\geq 0}$ is precompact in $\mathscr{C}^{1,\alpha}(-M,L),$ $\mbox{ for all } 0\leq\alpha <1$.
\end{enumerate}
\end{Lemma}
\begin{Proof}
The proof is completely analogous to the proof of Lemma \ref{Thm:properties_sequence_1d_scheme} and shall be omitted.
\end{Proof}
\begin{Proposition}[Existence and uniqueness, truncated $\mathcal{H}_{\infty}$-problem]\label{lem_uniqueness:H_infty}  Let $\Theta_j(\cdot,\cdot)$ be the sequence defined in \eqref{H_2D_existence_sequence_H_infty}.  Define $$\Theta_{(-M,L)}(x,y) = \inf_{j \in \mathbb{N}}\left(\Theta_j(x,y)\right).$$ Then $\Theta_{(-M,L)}(x,y)$ is the unique solution $ \Theta_{(-M,L)}(\cdot, \cdot)$ to \eqref{2D_H_infty_problem}.\end{Proposition}
\begin{Proof} 
Existence follows as in Proposition \ref{existence_lemma:1_leadsto_0}. The uniqueness proof differs slightly.  Let $w$  be a solution to  \eqref{2D_H_infty_problem}. Now generate a solution $\Theta_{(-M,L)}$ using the iteration scheme. We claim those two solutions coincide, thus proving uniqueness. We conclude from Lemma \ref{Thm:properties_sequence_H_infty}(i) that $\Theta_{(-M,L)} \geq w >0 $ in $\Omega_{(-M,L)}$. Now assume that $\Theta_{(-M,L)} \not\equiv w$
As both functions solve the PDE, we can compute
\begin{equation*}
\int_{\partial \Omega_{(-M,L)}}  \left(- \frac{\partial \Theta_{(-M,L)}}{\partial N} + \frac{\partial w}{\partial N}\right)g_{(-M,L)}dS+ \int_{ \Omega_{(-M,L)}}  \left[ (\Theta_{(-M,L)})^2 -  w^2\right] \Theta_{(-M,L)} w dxdy =0  \end{equation*}
where the first term was generated from integration by parts. We conclude from $\Theta_{(-M,L)} \geq w >0 $ that the first term is non-negative, while the second term is clearly strictly positive. This contradiction proves uniqueness.
\end{Proof}
 \subsection{Properties of solutions to the truncated problem}\label{s:32}

We want to study some properties of the mappings $(M,L) \mapsto \Theta_{(-M,L)}(\cdot,\cdot)$.

\begin{Lemma}[Comparison principles, $\mathcal{H}_{\infty}$-problem]\label{Thm:sub_super_solutions-H_infty} Let $\Theta_{(-M,L)}(\cdot,\cdot)$ be the solution from Proposition \ref{lem_uniqueness:H_infty}.
\noindent
\begin{enumerate}[label=(\roman*), ref=\theTheorem(\roman*)]
\hangindent\leftmargin
 \item \emph{(2D supersolutions)}\label{Thm:sub_super_solutions-H_infty:2D_supersolutions} 
 If $v$ satisfies, in the sense of distributions,
\begin{equation}\label{Thm:sub_super_solutions-H_infty:2D_supersolutions:supersolution_lemma_number_1-H_infty}
\Delta v  + \mu(x) v- v^3 \leq 0,\ (x,y)\in \Omega_{(-M,L)}, \quad \left(v - g_{(-M,L)}\right)\big|_{\partial \Omega_{(-M,L})} \geq 0, \quad   0 \leq v \leq 1,
\end{equation}
then $v \geq \Theta_{(-M,L)}$ in $\Omega_{(-M,L)}.$ In particular, $v \geq \Theta_{(-M,L)}$ in $\Omega_{(-M,L)}$ for any solution of 
\begin{equation*}
\Delta v  + \mu(x) v- v^3 =0,\ (x,y)\in \Omega_{(-M,L)}, \quad \left(v - g_{(-M,L)}\right)\big|_{\partial \Omega_{(-M,L})} \geq 0, \quad   0 \leq v \leq 1.
\end{equation*}

\item\emph{(2D subsolutions)}
If $v$ satisfies, in the sense of distributions,
\begin{equation}\label{Thm:sub_super_solutions-H_infty:2D_supersolutions:supersolution_lemma_number_2-H_infty}
\Delta v  + \mu(x) v- v^3 \geq 0,\ (x,y)\in \Omega_{(-M,L)}, \quad \left(v - g_{(-M,L)}\right)\big|_{\partial \Omega_{(-M,L})} \geq 0, \quad   0 \leq v \leq 1,
\end{equation}
then $v \leq \Theta_{(-M,L)}$ in $\Omega_{(-M,L)}.$ In particular, $v \leq \Theta_{(-M,L)}$ in $\Omega_{(-M,L)}$ for any solution of 
\begin{equation*}
\Delta v  + \mu(x) v- v^3 =0,\ (x,y)\in \Omega_{(-M,L)}, \quad \left(v - g_{(-M,L)}\right)\big|_{\partial \Omega_{(-M,L})} \leq 0, \quad   0 \leq v \leq 1.
\end{equation*}

\item 
The functions  from the one-dimensional problem, $v(y) = \theta_{(-M,0)}(y)  $  and $w(x) = \theta_{(-M,L)}(x)$, are supersolutions to \eqref{2D_H_infty_problem} in $\Omega_{(-M,L)}$. Consequently, for any $L, M >0$ we have that 
 \begin{eqnarray}\label{useful_for_monotonicity}
 \Theta_{(-M,L)}(x,y)\leq \min\{\theta_{(-M,L)}(x), \theta_{(-M,0)}(y)\}. 
 \end{eqnarray}
 \end{enumerate}
\end{Lemma}

\begin{Proof} Inequality \eqref{Thm:sub_super_solutions-H_infty:2D_supersolutions:supersolution_lemma_number_2-H_infty} is a direct consequence of inequality \eqref{Thm:sub_super_solutions-H_infty:2D_supersolutions:supersolution_lemma_number_1-H_infty}.
For the result in(i) now simply notice that the iterative method could be initialized with $\Theta_0(x,y) = v(x,y)$. A (possibly different) sequence $\left(\bar{U}_{j}\right)_{j \in \mathbb{N}}$ would be generated, where the property $v(x,y) \geq \bar{U}_j(x,y)$ can be verified through the same arguments used in the proof of Proposition \ref{Thm:properties_sequence_1d_scheme:prop_a}. By uniqueness, $\Theta_{(-M,L)}(x,y) = \inf_{j\in \mathbb{N}}\bar{U}_{j} \leq v(x,y)$. Part (ii) can be proven in a completely analogous fashion. To prove (iii), the result holds since $v$ and $w$ satisfy \eqref{Thm:sub_super_solutions-H_infty:2D_supersolutions:supersolution_lemma_number_1-H_infty}, hence both are supersolutions.
\end{Proof}
Following the ideas of the proof in one space dimension, we define extension operators in order to compare solutions for different values of $M,L$. Therefore, define, 
\begin{equation}
 \mathscr{E}\left[u\right](x,y) = \left \{\begin{array}{cc}
                          u(x,y), &   \quad  \mbox{for} \quad  (x,y) \in \Omega_{(-M,L)}\\
                           \mathscr{E}\left[\theta_{(-M,L)}\right](x)\cdot \mathscr{E}\left[\theta_{(-M,0)}\right](y),&\quad  \mbox{for} \quad (x,y) \in \mathbb{R}^2 \setminus \Omega_{(-M,L)}.
                          \end{array}\right. \label{extension_operator:H_infty}  \end{equation} 
Notice here that we use the same symbols for the one- and two-dimensional extension operators, slightly abusing notation, distinguishing between the two through the domain of definition of the function $\mathscr{E}$ is applied to.

\begin{Proposition}[Properties of the extension operator, $\mathcal{H}_{\infty}$-problem]\label{Thm:monotonicity_properties_H_infty_problem} The following properties of  $\mathscr{E}[\cdot]$ hold.
\begin{enumerate}[label=(\roman*), ref=\theTheorem(\roman*)]
\hangindent\leftmargin
 \item \emph{(Monotonicity of $\mathscr{E}$)}\label{Thm:monotonicity_properties_H_infty_problem:monotonicity_extension_operator}
  We have $0 \leq \mathscr{E}\left[\Theta_{(-M,L)}\right](\cdot, \cdot) \leq 1$. Furthermore, if $w$ is only defined in a subset $A \subset \mathbb{R}^2$ so that  $ \Omega_{(-M,L)} \subset A \subset \Omega_{(-\infty,L)}$, $0 \leq w(\cdot) \leq 1$ and $w(\cdot) \leq \Theta_{(-M,L)}(\cdot, \cdot)$,  then 
$0 \leq \mathscr{E}\left[w\right](\cdot) \leq \mathscr{E}\left[\Theta_{(-M,L)}(\cdot, \cdot)\right](\cdot)  \, \, \mbox{in} \, \, \mathbb{R}^2 $.                                                                                                                                                                                                                                        
 \item \emph{(Monotonicity in $M$)}\label{Thm:monotonicity_properties_H_infty_problem:2D_monotonicity_in_M}
Let  $0 \leq M <\widetilde{M}$ and $L \geq 0$ be fixed. Then $
M <\widetilde{M} \implies   \mathscr{E}\left[\Theta_{(-\widetilde{M},L)}\right](x,y) \leq  \mathscr{E}\left[\Theta_{(-M,L)}\right](x,y). $
 \item \emph{(Monotonicity in $L$)} \label{Thm:monotonicity_properties_H_infty_problem:2D_monotonicity_in_L}
Let  $0 \leq L <\widetilde{L}$ and $M \geq 0$ be fixed. Then $
L <  \widetilde{L} \implies   \mathscr{E}\left[\Theta_{(-M,L)}\right](x,y) \leq  \mathscr{E}\left[\Theta_{(-M,\widetilde{L})}\right](x,y). $
 \item \emph{(Monotonicity in $x$)}\label{Thm:monotonicity_properties_H_infty_problem:2D_monotonicity_in_x}
Let  $L,M,y$be fixed. Then
the mapping  $ x \to \mathscr{E}\left[\Theta_{(-M,L)}\right](x,y)$ is non-increasing. 

\item \emph{(Monotonicity in $y$)}\label{Thm:monotonicity_properties_H_infty_problem:2D_monotonicity_in_y}
Let  $L,M,x$be fixed. Then
the mapping  $ y \to \mathscr{E}\left[\Theta_{(-M,L)}\right](x,y)$ is non-decreasing. 
\end{enumerate}
\end{Proposition}

\begin{Proof}
Assertion (i) follows immediately from the definition of $\mathscr{E}$. To prove (ii), let  $M <\widetilde{M}$. According to (i), it suffices to compare 
\[
v(\cdot,\cdot)  := \mathscr{E}\left[\Theta_{(-\widetilde{M},L)}\right]\big|_{\mathcal{S}_{(-M,L)}}(\cdot,\cdot) = \Theta_{(-\widetilde{M},L)}\big|_{\mathcal{S}_{(-M,L)}}(\cdot,\cdot)
,\]
and $\Theta_{(-M,L)}(\cdot,\cdot)$ in $\Omega_{(-M,L)}$. One quickly verifies  that $v(\cdot, \cdot)$ is a subsolution to  \eqref{2D_H_infty_problem} in $\Omega_{(-M,L)},$ which proves the claim. Indeed, on the boundary of $\Omega_{(-M,L)}$ we have
 \begin{eqnarray}
v(x,y) = \left\{\begin{array}{ccc}
          \theta_{(-\widetilde{M},L)}(x)\theta_{(-\widetilde{M},0)}(0)   & = 0 & \mbox{for} \quad  x\in (-M,L), y = 0  \\
          \theta_{(-\widetilde{M},L)}(L)\theta_{(-\widetilde{M},0)}(y) & = 0  & \mbox{for} \quad  x = L, y \in (- M,0).  
         \end{array}  \right.
 \end{eqnarray}
On $x = -M$, $y \in (-M,0)$,  we have,
\begin{eqnarray*}
v(-M,y)\leq  \min\{\theta_{(-\widetilde{M},L)}(-M), \theta_{(-\widetilde{M},0)}(y)\} \leq \theta_{(-\widetilde{M},0)}(y) \leq \theta_{(-M,L)}(y) = g_{(-M,L)}(-M, y),  
\end{eqnarray*}
where the first inequality is a consequence of  \eqref{useful_for_monotonicity} and the last inequality follows from Proposition \ref{Thm:monotonicity_properties_1_to_0_1_to_0}(ii). A similar reasoning implies that $v(x,-M)\leq g_{(-M,L)}(x, -M)$ for  $x\in (-M,L),$ $y = -M$.
In conclusion,  $\left(v - g_{(-M,L)}\right)\big|_{\partial \Omega_{(-M,L)}} \leq 0$. As $v$ also solves the PDE in $\Omega_{(-M,L)}$ we can use Lemma \ref{Thm:sub_super_solutions-H_infty} to conclude the result. 

To show (iii), let  $L < \widetilde{L}  $. Using the monotonicity property (i), it suffices to compare $$v(\cdot,\cdot)  := \mathscr{E}\left[\Theta_{(-M,\widetilde{L})}\right]\big|_{\Omega_{(-M,L)}}(\cdot,\cdot) = \Theta_{(-M,\widetilde{L})}\big|_{\Omega_{(-M,L)}}(\cdot,\cdot)$$ and $\Theta_{(-M,L)}(\cdot,\cdot)$ in $\Omega_{(-M,L)}$. We will show  that $v(\cdot, \cdot)$ is a supersolution to  \eqref{2D_H_infty_problem} in $\Omega_{(-M,L)}$ by verifying the properties in \eqref{Thm:sub_super_solutions-H_infty:2D_supersolutions:supersolution_lemma_number_1-H_infty}, i.e., comparing these functions on the boundary of $\Omega_{(-M,L)}$. Indeed, $ v(x,y) =    \theta_{(-M,\widetilde{L})}(x)\theta_{(-M,0)}(0)    = 0 $  for  $x\in (-M,L)$, $y = 0.$ On  $x = -M$, $y \in (- M,0)$,  
\[
   v(-M,y) = \theta_{(-M,\widetilde{L})}(-M)\theta_{(-M,0)}(y)  =\theta_{(-M,0)}(y) = g_{(-M,\widetilde{L})}(-M,y) = \Theta_{(-M,\widetilde{L})}(-M,y)
.\]
On  $ x = L$, $y \in (-M,0)$, simply notice that $v(L,y) \geq 0 = \Theta_{(-M,L)}(L,y).$
On  $x\in (-M,L)$, $y = -M $,
\begin{align*}
v(x,-M) &= \theta_{(-M,\widetilde{L})}(x)\theta_{(-M,0)}(-M) =  \theta_{(-M,\widetilde{L})}(x) \\
&\stackrel{\eqref{useful_for_monotonicity}}{\geq} \theta_{(-M,L)}(x) = g_{(-M,L)}(x,-M) = \Theta_{(-M,L)}(x,-M). \end{align*}
Since $v$ also solves the PDE in $\Omega_{(-M,L)}$ we can use Lemma \ref{Thm:sub_super_solutions-H_infty}(i)  to conclude the result. 

To prove (iv), let $x_0>0$ and $M,L> x_0$. The inequality follows if we show that 
\begin{eqnarray}\label{monotonicity_in_x_ineq_simpler:H_infty}
 \Theta_{(-M,L)}(x + x_0,y) \leq \Theta_{(-M,L-x_0)}(x,y) \quad  \mbox{for} \quad (x,y) \in  \Omega_{(-M,L-x_0)}. 
\end{eqnarray}
Indeed, if inequality \eqref{monotonicity_in_x_ineq_simpler:H_infty} holds then 
$$\Theta_{(-M,L)}(x +x_0,y) \leq \Theta_{(-M,L-x_0)}(x,y) \leq  \Theta_{(-M,L)}(x,y) $$
 for $(x,y) \in  \Omega_{(-M,L-x_0)}$, using (iii) in the last inequality. Notice that the inequality in (iii) holds without the extension operator, since both $\Theta_{(-M,L)}(\cdot,\cdot)$ and $\Theta_{(-M,L)}(\cdot + x_0,\cdot)$ agree with their extensions in $\Omega_{(-M,L-x_0)}$. Since moreover $\mathscr{E}\left[\Theta_{(-M,L)}(x +x_0,y)\right] =0$ for all $x\geq L-x_0$ we conclude that 
\[
\mathscr{E}\left[\Theta_{(-M,L)}\right]\big|_{\Omega_{(-M,L)}}(x +x_0,y) \leq \Theta_{(-M,L)}(x,y),
\]
for $(x,y) \in  \Omega_{(-M,L)}$. Monotonicity, (iv), now follows after applying the extension operator and using its monotonicity properties.
 
To show \eqref{monotonicity_in_x_ineq_simpler:H_infty}, we let $w(\cdot, \cdot):= \Theta_{(-M,L)}(\cdot + x_0,\cdot)$, defined in $\Omega_{(-M,L-x_0)}$, and show that $w(\cdot,\cdot)$ is a subsolution to \eqref{2D_H_infty_problem} in $\Omega_{(-M,L-x_0)}.$ For that, we need to check the relevant properties form Lemma \ref{Thm:sub_super_solutions-H_infty}. Indeed, we have
  \begin{align*}
w(-M,y) = \Theta_{(-M,L)}(-M + x_0,y ) \stackrel{\eqref{useful_for_monotonicity}}{\leq} &\min\{\Theta_{(-M,L-x_0)}(-M+x_0),\Theta_{(-M,0)}(y)\} \nonumber\\
                                        \leq &\theta_{(-M,0)}(y)   = g_{(-M,L-x_0)}(-M,y) =  \Theta_{(-M,L-x_0)}(-M,y)
  \end{align*}
for $x = -M,$ $y \in [-M,0]$. Furthermore, $w(x,0) = \Theta_{(-M,L)}(x + x_0, 0 ) =0 = \Theta_{(-M,L-x_0)}(x,0)$ for $y=0$, $x \in [-M,L-x_0]$ and $ w(L-x_0,y) = \Theta_{(-M,L)}(L,y ) = 0  \leq   \Theta_{(-M,L-x_0)}(L-x_0,y)$ for $x = L - x_0$, $y \in [-M,0].$ 
Lastly, 
 \begin{align*}
w(x,-M) &= \Theta_{(-M,L)}(x + x_0,-M ) =g_{(-M,L)}(x + x_0,-M ) = \Theta_{(-M,L)}(x+x_0) \nonumber\\
                                       & \leq \theta_{(-M,L)}(x) =  \Theta_{(-M,L-x_0)}(x,-M),
  \end{align*}  
for $y=-M$, $x \in [-M,L-x_0],$ where we used Lemma \ref{Thm:monotonicity_properties_1_to_0_1_to_0}(iv) in the inequality.  It remains to show that  $\Delta w + \mu(x) w - w^3 \geq 0,$ is satisfied in the sense of distributions; this result follows using the same reasoning as in the proof of  Lemma \ref{Thm:monotonicity_properties_1_to_0_1_to_0}(iv).

The proof of (v) is similar and will  be omitted. 

\end{Proof}
We are now ready to pass to the limit $M=\infty$. Define 
\begin{equation}
\Theta_{(-\infty,L)}(x,y) := \inf_{M >0 } \mathscr{E}\left[\Theta_{(-M,L)}\right](x,y) =\lim_{M \to +\infty } \mathscr{E}\left[\Theta_{(-M,L)}\right](x,y),\label{definition:theta_infty_L:H_infty} 
\end{equation}
where the last equality holds due to monotonicity of the mapping $M \mapsto \Theta_{(-M,L)}(x,y)$, Proposition \ref{Thm:monotonicity_properties_H_infty_problem}(ii).

%
%

\subsection{Properties of semi-truncated solutions}\label{properties_theta_minusinfty_L}
 In this section we verify monotonicity properties and limits at spatial infinity of the limits $\Theta_{(-\infty,L)}(x,y)$ constructed in \eqref{definition:theta_infty_L:H_infty}.

 \begin{Lemma}[Monotonicity of $\Theta_{(-\infty,L)}(x,y)$]\label{H_infty_L_problem_solution}
 The limits  $\Theta_{(-\infty,L)}(\cdot,\cdot)$ from \eqref{definition:theta_infty_L:H_infty} solve equation \eqref{2D_H_infty_problem} in  $\Omega_{(-\infty,L)}:= \{(x,y) \in \mathbb{R}^2 | x <L, y < 0 \}$. Furthermore, $\Theta_{(-\infty,L)}(x,y)$ is non-increasing in $x$ and $y$ and non-decreasing in $L$. 
\end{Lemma}
\begin{Proof} Pointwise convergence and uniform boundedness imply that the PDE is satisfied in the sense of distributions. Monotonicity follows as in the one-dimensional case from monotonicity properties of extension operators and monotonicity in the truncated problem after taking infima. 
\end{Proof}

\begin{Lemma}\label{2D_supersolutions_infty_L} Let $\theta_{(-M,L)}(\cdot)$ be a solution to the truncated $1 \leadsto 0$-problem in $(-M,L)$. Define $v(y) := \theta_{(-\infty,0)}(y)  $  and $w(x) := \theta_{(-\infty,L)}(x)$. Then $v(\cdot)$ and $w(\cdot)$ are supersolutions to  \eqref{2D_H_infty_problem} in $\Omega_{(-\infty,L)}$. Consequently, for any $L >0$ we have that 
 \begin{eqnarray*}
 \Theta_{(-\infty,L)}(x,y)\leq \min\{\theta_{(-\infty,L)}(x), \theta_{(-\infty,0)}(y)\}. 
 \end{eqnarray*}
 In particular, we have  
 \[
 \sup_{L>0}\left(\Theta_{(-\infty,L)}(x,y) \right)\leq \min\{\theta(x), \theta_{(-\infty,0)}(y)\},
 \]
 where $\theta=\theta_{(-\infty,\infty)}$.
 \end{Lemma}
\begin{Proof}The first assertion is direct consequence of Lemma \ref{Thm:sub_super_solutions-H_infty}(iii) and monotonicity  of $\Theta_{(-M,L)}(\cdot)$ and $\Theta_{(-\infty,L)}(\cdot)$ in its arguments. The second claim follows by passing to the limit $L=\infty$ from the fact that $\theta_{-\infty,L)}$ is increasing in $L$. 
 \end{Proof}
 
\begin{Proposition}[Asymptotics to 1D profiles, $\mathcal{H}_{\infty}$-problem]\label{H_infty_L_asymptotics_to_profiles}
 The semi-truncated solutions have  limits 
\begin{eqnarray*}
 \theta_{(-\infty,L)}(x) = \lim_{ y \to -\infty} \Theta_{(-\infty,L)}(x,y) \quad \mbox{and} \quad \theta_{(-\infty,0)}(y)= \lim_{ x \to -\infty} \Theta_{(-\infty,L)}(x,y)\,
\end{eqnarray*}
\end{Proposition}
For the proof, we will need two auxiliary lemmas. 

\begin{Lemma} \label{Thm:subsolutions_alpha_beta:part_1}
Choose $\alpha,\beta $ with $\frac{1}{\alpha^2} + \frac{1}{\beta^2} =1$, $L\geq 0$, and let
$ w_{(\alpha, \beta)}(x,y) := \theta_{(-\infty,L)}({x}/\alpha)\cdot \theta_{(-\infty,0)}({y}/\beta).$
Then 
\[
\Delta w_{(\alpha, \beta)} + \mu(x) w_{(\alpha,\beta)} - w_{(\alpha,\beta)}^3 \geq 0,
\]
in the sense of distributions in $\Omega_{(-\infty,\alpha\cdot L)}$ (and, consequently, in $\Omega_{(-\infty, L)} \subset \Omega_{(-\infty,\alpha\cdot L)}$). 
\end{Lemma}

\begin{Proof}
The proof is a direct calculation. We  drop indices for $w$ and $\theta$ and find
\begin{align*}
 \Delta w(x,y) &= \theta\left(\frac{y}{\alpha}\right)\partial_x^2\left[\theta\left(\frac{x}{\beta}\right)\right] + \theta\left(\frac{x}{\beta}\right)\partial_y^2\left[ \theta\left(\frac{y}{\alpha}\right)\right]  = \nonumber\\
 &=\frac{1}{\beta^2}\theta\left(\frac{y}{\alpha}\right)\cdot \left[ - \mu\theta\left(\frac{x}{\beta}\right)+ \theta^3\left(\frac{x}{\beta}\right)  \right] + \frac{1}{\alpha^2}\theta\left(\frac{x}{\beta}\right)\cdot \left[ - \mu\theta\left(\frac{y}{\alpha}\right) + \theta^3\left(\frac{y}{\alpha}\right)  \right] = \nonumber\\
& = -  \mu\theta\left(\frac{x}{\beta}\right)\theta\left(\frac{y}{\alpha}\right) +\left[ \frac{1}{\beta^2}\theta^3\left(\frac{x}{\beta}\right)\theta\left(\frac{y}{\alpha}\right) + \frac{1}{\alpha^2}\theta\left(\frac{x}{\beta}\right)\theta^3\left(\frac{y}{\alpha}\right)\right] \geq \nonumber\\
& \geq -  \mu\theta\left(\frac{x}{\beta}\right)\theta\left(\frac{y}{\alpha}\right)  + \left[\theta\left(\frac{x}{\beta}\right)\theta\left(\frac{y}{\alpha}\right)\right]^3,
 \end{align*}
which proves the lemma.
\end{Proof}
\begin{Lemma} \label{Thm:subsolutions_alpha_beta:part_2}
 Let $\widetilde{L} \leq L$. Define  $w_{(\alpha, \beta)}(x,y) := \theta_{(-\infty,\widetilde{L})}(x/\alpha)\cdot \theta_{(-\infty,0)}(y/\beta)$  for  $\alpha =L/\widetilde{L}$ and $\beta = L/\sqrt{ L^2 -\widetilde{L}^2 }.$  Then 
$$w_{(\alpha, \beta)}(x,y) \leq u_{(-\infty,L)}(x,y) \quad \mbox{in} \quad \Omega_{(-\infty,\alpha \widetilde{L})} =\Omega_{(-\infty, L)}.$$
 Furthermore if $L =0$ then we can  choose $\alpha$ and $\beta$ arbitrarily subject to  $\frac{1}{\alpha^2} + \frac{1}{\beta^2} =1$.
\end{Lemma}

 \begin{Proof} It is clear that $\frac{1}{\alpha^2}+ \frac{1}{\beta^2} =1$. The result follows once we show that 
\begin{eqnarray}\label{simpler_result:H_infty_problem}
w(x,y) \leq \Theta_{(-M,L)}(x,y)  \quad \mbox{for}  \quad (x,y) \in  \Omega_{(-M, L)}.
\end{eqnarray}
Indeed, from  inequality \eqref{simpler_result:H_infty_problem} and the monotonicity of the extension operator (Proposition \ref{Thm:monotonicity_properties_H_infty_problem:monotonicity_extension_operator}) one can conclude that $w(x,y) \leq \mathscr{E}\left[\Theta_{(-M,L)}\right](x,y)$ for $(x,y) \in  \Omega_{(-\infty, L)}$ and now take the infimum in $M$ on the right hand side. To prove \eqref{simpler_result:H_infty_problem}  we show that $w(\cdot, \cdot)$ is a subsolution to  \ref{2D_H_infty_problem} in $\Omega_{(-M, L)}.$ Lemma \ref{Thm:subsolutions_alpha_beta:part_1} establishes the first condition of Lemma   \ref{Thm:sub_super_solutions-H_infty} and we can focus on  the boundary. For $x = L$, $y \in (-M,L),$ 
 \begin{align*}
[\Theta_{(-M,L)} - w](L,y)  &=  g_{(-M,L)}(L,y) - \theta_{(-M,\widetilde{L})}\left(\frac{L \cdot \widetilde{L}}{L}\right)\theta_{(-M,0)}\left(\frac{y}{\beta}\right)\\  
                              &=\theta_{(-M,L)}(L)\theta_{(-M,0)}(y)  - \theta_{(-M,\widetilde{L})}\left(\frac{L \cdot \widetilde{L}}{L}\right)\theta_{(-M,0)}\left(\frac{y}{\beta}\right)  =0
 \end{align*}
On $x \in (-M,L)$, $y = 0,$  we have  $\Theta_{(-M,L)}(x,0) - w(x,0) = 0.$ On $x = -M$, $y \in (-M,L),$
 \begin{align*}
[\Theta_{(-M,L)} - w](-M,y) & = g_{(-M,L)}(-M,y) - w(-M,y) = \\
 & \stackrel{\text{Def.}}{=}\theta_{(-M,L)}(-M)\theta_{(-M,0)}(y) - \theta_{(-M,\widetilde{L})}\left(\frac{-M\cdot\widetilde{L} }{L}\right)\theta_{(-M,0)}\left(\frac{y}{\beta}\right)  \\
 & \geq \theta_{(-M,0)}(y) - \theta_{(-M,0)}\left(\frac{y}{\beta}\right) \stackrel{\beta \geq1.}{\geq} 0
 \end{align*}
where the last inequality is due to the fact that the mapping $y \mapsto \theta_{(-M,0)}(y) $ is non-increasing Proposition \ref{1D_monotonicity_infty_L}.
On $x \in (-M,L)$, $y = -M,$
 \begin{align*}
[\Theta_{(-M,L)} - w](x,-M) & = g_{(-M,L)}(x,-M) - w(x,-M) = \\
 & \stackrel{\text{Def.}}{=} \theta_{(-M,L)}(x)\theta_{(-M,0)}(-M) - \theta_{(-M,\widetilde{L})}\left(\frac{x\widetilde{L}}{L}\right)\theta_{(-M,0)}\left(\frac{-M}{\beta}\right)  \\
 & \geq \theta_{(-M,L)}(x) - \theta_{(-M,\widetilde{L})}\left(\frac{x\widetilde{L}}{L}\right)  \\
 & {\geq} \theta_{(-M,\widetilde{L})}(x) - \theta_{(-M,\widetilde{L})}\left(\frac{x\widetilde{L}}{L}\right)  \stackrel{L\geq \widetilde{L}}{\geq} 0.
 \end{align*}
Here, the second last inequality is due to Proposition \ref{Thm:monotonicity_properties_H_infty_problem}(iii) and the last inequality follows from the fact that the mapping $x \mapsto \theta_{(-M,\widetilde{L})}(x)$ is non-increasing, Proposition \ref{Thm:monotonicity_properties_1_to_0_1_to_0}(iv).
Hence,  $\left(g_{(-M,L)}(x,y)  - w(x,y)\right)\big|_{\partial \Omega_{(-M, L)} }\geq 0. $ Using the maximum principle, we conclude $w(x,y)\leq \Theta_{(-M,L)}(x,y)$  in $\Omega_{(-M, L)}$, which finishes the proof of \eqref{simpler_result:H_infty_problem}. The proof in the case $L =0$ is analogous.
\end{Proof}

The next lemma recovers a result of Dang, Fife and Peletier \cite[Lemma 4]{dang1992saddle}.
\begin{Lemma}[Uniform convergence in compact sets]\label{lemma_uniform_convergence_compact_sets} The limits
\begin{eqnarray*}
 H_L(x) := \lim_{ y \to -\infty} \Theta_{(-\infty,L)}(x,y) \quad \mbox{and} \quad V_L(y) := \lim_{ x \to -\infty} \Theta_{(-\infty,L)}(x,y)
\end{eqnarray*}
are attained uniformly in compact subsets of $(-\infty,0]$ and $\mathbb{R}$, respectively.
\end{Lemma}
\begin{Proof}
 Fix $L$, which we assume to be positive (the case $L=0$ has a similar proof). Given  $K_1 \subset \mathbb{R}$ and $K_2 \subset (-\infty,0]$  nonempty compact sets and  $\epsilon >0$. We firstprove the result for $H_L(\cdot)$. Since $K_1$ is compact and $\theta_{(-\infty,L)}(\cdot)$ is continuous, hence uniformly continuous over $K_1$, we can choose  $\delta_1>0$ such that 
 \begin{eqnarray}\label{unif_convergence_1}
\sup_{|x -y| \leq \delta_1}\left|\theta_{(-\infty,L)}(x) -\theta_{(-\infty,L)}(y) \right| \leq \frac{\epsilon}{3}, \quad \mbox{whenever} \quad x, y \in  K_1.
 \end{eqnarray}
 Since $L \mapsto \theta_{(-\infty,L)}(\cdot) $ is continuous in the sup norm, Lemma \ref{1D_lemma_continuity_in_L_M},  we can find  $\delta_2>0$ such that
 \begin{eqnarray}\label{unif_convergence_2}
|L - \widetilde{L}|< \delta_2 \implies \sup_{x \in \mathbb{R}}\left|\theta_{(-\infty,L)}(x) -\theta_{(-\infty,\widetilde{L})}(x) \right| \leq \frac{\epsilon}{3}.  
 \end{eqnarray}
 Further, we can choose a $\delta_3 >0$ such that 
 \begin{eqnarray}\label{unif_convergence_3}
|L - \widetilde{L}|< \delta_3 \implies \left(\sup_{x \in K_1} |x|\right)\cdot \left|\frac{L - \widetilde{L}}{\widetilde{L}} \right| \leq \delta_1
 \end{eqnarray}
 Now choose $\widetilde{L}$ such that $|L - \widetilde{L}|\leq \min\{\delta_1, \delta_2,\delta_3\}$ and $\alpha:= {\widetilde{L}}/{L} >1$. Now, for the associated and fixed $\beta = \beta(\widetilde{L})$, take $C>0$ sufficiently large such that 
 \begin{equation}\label{unif_convergence_4}
 \theta_{(-\infty,0)}\left(\frac{y}{\beta} \right) \geq 1 - \frac{\epsilon}{3}   \quad \mbox{whenever} \quad y \leq -C
 \end{equation}
Hence, using first Lemma \ref{2D_supersolutions_infty_L} and subsequently Lemma \ref{Thm:subsolutions_alpha_beta:part_2},
 \begin{align*}
 0 &{\leq}\theta_{(-\infty, L)}(x) -  \Theta_{(-\infty,L)}(x,y) {\leq} \theta_{(-\infty,L)}(x) - \theta_{(-\infty, \widetilde{L})}\left(\frac{x}{\alpha}\right)\theta_{(-\infty, 0)}\left(\frac{y}{\beta}\right) \\
 & \quad \leq \left\{\theta_{(-\infty,L)}(x) -\theta_{(-\infty,L)}\left(\frac{x}{\alpha} \right)\right\} + [\theta_{(-\infty,L)} -  \theta_{(-\infty, \widetilde{L})}]\left(\frac{x}{\alpha}\right) \\
 &\quad \quad +  \theta_{(-\infty, \widetilde{L})}\left(\frac{x}{\alpha}\right)\left[ 1 - \theta_{(-\infty, 0)}\left(\frac{y}{\beta}\right)\right]  = I_1 + I_2 + I_3 \leq \epsilon,
 \end{align*}
where $I_1 \leq\frac{\epsilon}{3}$ due to $\eqref{unif_convergence_1}$ and $\eqref{unif_convergence_3}$; $I_2  \leq\frac{\epsilon}{3}$ due to $\eqref{unif_convergence_2}$; $I_3 \leq\frac{\epsilon}{3}$ due to $\eqref{unif_convergence_4}$ and the fact that $\theta(\cdot) \leq 1$.  

We now turn to the horizontal limit $V_L(\cdot)$. Since $K_2 \subset (-\infty,0]$ is compact and $\theta_{(-\infty,0)}(\cdot)$ is continuous, hence uniformly continuous over $K_2$, one can choose a $\delta_4>0$ such that 
 \begin{eqnarray*}
\sup_{|y_1 -y_2| \leq \delta_4}\left|\theta_{(-\infty,0)}(y_1) -\theta_{(-\infty,0)}(y_2) \right| \leq \frac{\epsilon}{2},\quad \mbox{whenever} \quad y_1, y_2 \in  K_2.
 \end{eqnarray*}
Next, we can choose an $M>0$ sufficiently large such that 
 \begin{eqnarray*}
\widetilde{L} > M \implies \left(\sup_{y \in K_2} |y|\right)\cdot \left|1 - \frac{\sqrt{\widetilde{L}^2 - L^2}}{\widetilde{L}} \right| \leq \delta_4.
 \end{eqnarray*}
 Now, pick $C>0$ sufficiently large such that $ \theta_{(-\infty,\widetilde{L})}\left(\frac{x}{\alpha} \right) \geq 1 - \frac{\epsilon}{2}$ whenever $x \leq -C$. Hence, using again first Lemma \ref{2D_supersolutions_infty_L} and subsequently Lemma \ref{Thm:subsolutions_alpha_beta:part_2},
 \begin{align*}
 0 \leq&\theta_{(-\infty, 0)}(y) -  \Theta_{(-\infty,L)}(x,y)\\
 {\leq} &\theta_{(-\infty,0)}(y) - \theta_{(-\infty, \widetilde{L})}\left(\frac{x}{\alpha}\right)\theta_{(-\infty, 0)}\left(\frac{y}{\beta}\right) \\
 \leq& \left\{\theta_{(-\infty,0)}(y) -\theta_{(-\infty,0)}\left(\frac{y}{\beta} \right)\right\} + \theta_{(-\infty, 0)}\left(\frac{y}{\beta}\right)\left[ 1 -  \theta_{(-\infty, \widetilde{L})}\left(\frac{x}{\alpha} \right)\right] \leq \epsilon.
 \end{align*}
\end{Proof}

\begin{Proof}[of Proposition ~\ref{H_infty_L_asymptotics_to_profiles}] Boundedness and monotonicity properties of the functions $\Theta_{(-\infty,L)}(x,y)$ in  both $x$ and $y$ imply that horizontal and vertical limits  are well defined. Assume $L >0$.  From the previous lemma it suffices to show that $ \theta_{(-\infty,L)}(x)= \displaystyle{\lim_{ y \to -\infty}} \Theta_{(-\infty,L)}(x,y)$ holds for all $x$ large and $\theta_{(-\infty,0)}(y) := \displaystyle{\lim_{ x \to -\infty}} \Theta_{(-\infty,L)}(x,y)$ holds for all $y$ large. We first prove vertical $y$-limits.  Let   $\epsilon >0$ be given and $\widetilde{L}$ be chosen as in the previous lemma, $\alpha$ and $\beta$ fixed. Fix $C>0$ sufficiently large to that 
\begin{eqnarray}\label{convergence_inequalities_large_x}
\theta_{(-\infty, L)}\left(\frac{x}{\alpha}\right) \geq 1 - \frac{\epsilon}{3} \quad \mbox{and} \quad \theta_{(-\infty, \widetilde{L})}\left(\frac{x}{\alpha}\right) \geq 1 - \frac{\epsilon}{3}\quad \mbox{whenever} \quad x \leq -C. 
\end{eqnarray}
Finally, choose $C_2>0$ sufficiently large so that $\theta_{(-\infty, 0)}\left(\frac{y}{\beta}\right) \geq 1 - \frac{\epsilon}{3}.$ We conclude that, for all $x \leq C$, $y \leq C_2$,
\begin{align*}
 0 &\leq\theta_{(-\infty, L)}(x) -  \Theta_{(-\infty,L)}(x,y)  \\
 &\leq \theta_{(-\infty,L)}(x) - \theta_{(-\infty, \widetilde{L})}\left(\frac{x}{\alpha}\right)\theta_{(-\infty, 0)}\left(\frac{y}{\beta}\right) \\
 & \leq \left\{\theta_{(-\infty,L)}(x) -\theta_{(-\infty,L)}\left(\frac{x}{\alpha} \right)\right\} + [\theta_{(-\infty,L)} -  \theta_{(-\infty, \widetilde{L})}]\left(\frac{x}{\alpha}\right)  +  \theta_{(-\infty, \widetilde{L})}\left(\frac{x}{\alpha}\right)\left[ 1 - \theta_{(-\infty, 0)}\left(\frac{y}{\beta}\right)\right] \\
 & = I_1 + I_2 + I_3 \leq \epsilon.
 \end{align*}
We next turn to horizontal $x$-limits. Fix $\widetilde{L} >0$ sufficiently large so that the inequalities \eqref{convergence_inequalities_large_x} hold. Pick $C_3>0$ sufficiently large so that $ \theta_{(-\infty,0)}\left(\frac{y}{\beta} \right) \geq 1 - \frac{\epsilon}{2}$ whenever $y \leq -C_3.$ For all $x \leq -C_3$ and $y \leq C$  we obtain
 \begin{align*}
 0 &\leq\theta_{(-\infty, 0)}(y) -  \Theta_{(-\infty,L)}(x,y) \\
 &\leq \theta_{(-\infty,0)}(y) - \theta_{(-\infty, \widetilde{L})}\left(\frac{x}{\alpha}\right)\theta_{(-\infty, 0)}\left(\frac{y}{\beta}\right) \\
 & \leq \left\{\theta_{(-\infty,0)}(y) -\theta_{(-\infty,0)}\left(\frac{y}{\beta} \right)\right\} + \theta_{(-\infty, 0)}\left(\frac{y}{\beta}\right)\left[ 1 -  \theta_{(-\infty, \widetilde{L})}\left(\frac{x}{\alpha} \right)\right] \leq \epsilon. 
 \end{align*}
 \end{Proof}
A weaker version of Lemma \ref{Thm:subsolutions_alpha_beta:part_1} was proven in \cite{schatzman1995stability}, stating in our notation that
$$\theta_{(-M,0)}\left(\frac{x}{\sqrt{2}}\right)\cdot \theta_{(-M,0)}\left(\frac{y}{\sqrt{2}}\right) \leq \Theta_{(-\infty,0)}(x,y),$$
in  $\Omega_{(-\infty,0)}$. Using this, one can show that 
$$\theta_{(-M,0)}(x) \leq \lim_{y \to -\infty}\Theta_{(-\infty,L)}(x,y)  \leq \theta_{(-M,L)}(x),$$
which is not sufficient to conclude that the last inequality is actually an equality. in this sense, our lemma strengthens the result in  \cite{schatzman1995stability}, implying uniform convergence on compact sets, Lemma \ref{lemma_uniform_convergence_compact_sets}.
%

\subsection{Passing to the limit}\label{H_infty_problem_solution}
We conclude the proof of  Theorem \ref{main_theorem:c_zero:H_infty}, $\kappa=\infty$ case, by passing to the limit $L=\infty$. Define 
\begin{eqnarray*}
  \Theta(x,y) = \sup_{L >0} \Theta_{(-\infty,L)}(x,y) =  \lim_{L\to \infty}\Theta_{(-\infty,L)}(x,y). 
 \end{eqnarray*}
We then have the following convergence result, which implies Theorem \ref{main_theorem:c_zero:H_infty}.
\begin{Proposition}\label{Thm:H_infty_problem_solution} The limit $\Theta(x,y)$ exists and 
 solves  \eqref{e:hinf}. Moreover, we have the pointwise  limits
 \begin{eqnarray*}
 \theta(x)) := \lim_{ y \to -\infty} \Theta(x,y) \quad \mbox{and} \quad \theta_{(-\infty,0)}(y) := \lim_{ x \to -\infty} \Theta(x,y),
\end{eqnarray*}
and $\Theta(x,y)$ is non-increasing in $x$ and in $y$.
\end{Proposition}

 \begin{Proof} Convergence to a solution in the sense of distributions follows as in Proposition \ref{existence_lemma:1_leadsto_0} from pointwise convergence and uniform boundedness. Monotonicity follows from monotonicity of the semi-truncated solutions taking monotone limits. We need to establish horizontal and vertical pointwise limits. 
 
Boundedness and monotonicity properties of the function $\Theta_{(-\infty, L)}$ in  both $x$ and $y$ imply that horizontal and vertical limits exist. 
By construction, and by Lemma  \ref{2D_supersolutions_infty_L},
\begin{eqnarray}\label{horizontal_and_vertical_limits_ineq_proof}
 \Theta_{(-\infty, L)}(x,y) \leq \Theta(x,y){\leq}  \min\{\theta(x), \theta_{(-\infty,0)}(y)\}. 
\end{eqnarray}
Taking the limit $y\to -\infty$, we find, using Proposition \ref{H_infty_L_asymptotics_to_profiles},
\begin{eqnarray*}
\theta_{(-\infty,L)}(x) \leq \liminf_{y \to -\infty} u(x,y)\leq \limsup_{y \to -\infty} u(x,y)\leq \theta(x)\end{eqnarray*}
On the left hand side we can let  $L \to \infty$ and invoke monotonicity of the mapping $L \mapsto \theta_{(-\infty,L)}(x) $  to conclude that the vertical limit is $\theta(x)$. Analogously, taking $x \to -\infty$ in inequality \eqref{horizontal_and_vertical_limits_ineq_proof}  gives, using again  Proposition \ref{H_infty_L_asymptotics_to_profiles},
\begin{eqnarray*}
\theta_{(-\infty,0)}(y) \leq \liminf_{x \to -\infty} u(x,y)\leq \limsup_{x \to -\infty} u(x,y)\leq \theta_{(-\infty,0)}(y).
\end{eqnarray*}
We finish the proof taking the limit $L \to \infty$. We shall prove exponential convergence towards the limiting profiles as $x\to\pm\infty$ or $y\to\infty$, in Corollary \ref{c:expinf}.

\end{Proof}


%
\section{Two-dimensional quenched patterns at zero speed --- periodic horizontal interfaces}\label{section:H_problem}
In this section, we will prove Theorem \ref{main_theorem:c_zero:H_infty} in the case $\pi<\kappa<\infty$. Much of the analysis here is similar, in fact slightly easier, than in the case $\kappa=\infty$. A key difference is the construction of subsolutions guaranteeing the convergence for $x\to-\infty$. Crucial to our approach is again the fact that the nonlinearity is odd. We solve our equation\eqref{intro_eq_1D_problem} in a strip $
 \mathcal{S}_{(-M,L)} := (-M,L)\times(0,\kappa)$, with  homogeneous Dirichlet boundary conditions at $y\in\{0,\kappa\}$. As in the previous section, we can continue the solution to a $y$-periodic solution by reflecting $u(x,-y)=-u(x,y)$ and $u(x,\kappa+y)=-u(\kappa-y)$. 

 \subsection{The truncated problem --- existence and properties of solutions}\label{s:4.1}
 Recall that we wish to find solutions $\Xi$ to $\Delta u +\mu(x) u - u^3=0$ in $\mathcal{S}_{(-\infty,\infty)}=\R\times (0,\kappa)$, with Dirichlet boundary conditions on $\R\times \{0,\kappa\}$, and asymptotics $u(x,y)\to 0$ for $x\to +\infty$, $u(x,y)-\bar{u}(y\to 0$ for $x\to -\infty$, where $\bar{u}(y)$ is the   $2\kappa$-periodic solutions of the one-dimensional problem. For the truncation, we conveniently define $  h_{(-M,L)}(x,y) := \theta_{(-M,L)}(x)\cdot\bar{u}(y)$. The truncated \emph{$\mathcal{H}$-problem} (with parameter $\kappa>\pi$), is 
  \begin{align}
\Delta u + \mu(x) u - u^3&=0, & (x,y)\in\mathcal{S}_{(-M,L)} ,\nonumber\\
u-h_{(-M,L)}& =0, & (x,y)\in\mathcal{S}_{(-M,L)},\label{H_2D_stripe_problem}
\end{align}
Throughout, we will suppress the dependence of $\Xi$ on $\kappa$ for ease of notation.
We again find uniform a priori bounds in $\mathscr{C}^{(1,\alpha)}(\overline{\mathcal{S}_{(-M,L)}}), $ exploiting that $0\leq u\leq 1$. 
 
Following the strategy in the $\mathcal{H}_\infty$ case, we define the iteration scheme
\begin{eqnarray}\label{H_2D_stripe_existence_sequence}%
\left\{\begin{array}{c}
  -  \Delta \Xi_{n+1}+ 5 \Xi_{n+1} =  (5 +\mu(x)) \Xi_n - \Xi_n^3  \\
 (\Xi_{n+1} - h_{-M,L})\big|_{\partial \mathcal{S}_{(-M,L)}}=0
\end{array}\right.
\end{eqnarray}
where $\Xi_0(\cdot)$ is chosen in the class 
\begin{align*}
 \Psi_{\mathcal{H}_{\infty}} &:= \left\{\Xi \in \mathscr{C}^{(1,\alpha)}(\mathcal{S}_{(-M,L)})\,\middle|\,(\Xi_{0} - g_L)\big|_{\partial \mathcal{S}_L} \geq 0,\ \Xi(x)\in[0,1] \mbox{ for all } x,y \right\}\\
 & \quad \cap \{  \Delta\Xi + \mu(x)\Xi- \Xi^3 \leq 0, \, \  \Delta\Xi + \mu(x)\Xi - \Xi^3 \not\equiv 0 \, \mbox{in the sense of distributions}\}.
\end{align*}
%
%
%
%
Following the arguments from Section \ref{section:H_infty}, we find unique solutions $\Xi_{(-M,L)}$ to the $\mathcal{H}$-problem through iteration of \eqref{H_2D_stripe_existence_sequence}. 

\subsection{Passing to the limit}

In order to pass to the limit, we introduce the extension operator $\mathscr{E}$ acting on functions defined on $\mathcal{S}_{(-M,L)}$, through
\begin{eqnarray*}
 \mathscr{E}\left[u\right](x,y) = \left \{\begin{array}{cc}
                          u(x,y), & \quad  \mbox{for} \quad  (x,y) \in \mathcal{S}_{(-M,L)},\\
                           \mathscr{E}\left[\Theta_{(-M,L)}\right](x)\cdot \bar{u}(y),&\quad  \mbox{for} \quad (x,y) \in \mathbb{R}\times[0,\kappa] \setminus \mathcal{S}_{(-M,L)}.
                          \end{array}\right. \end{eqnarray*} 
One then readily establishes comparison principles and monotonicity properties of $\mathscr{E}$, analogous to Lemma \ref{Thm:sub_super_solutions-H_infty} and Proposition \ref{Thm:monotonicity_properties_H_infty_problem}, respectively\footnote{Monotonicity in $y$ is restricted to $y\in [0,\kappa/2]$.}. We  subsequently pass to the limit $M\to\infty$, finding solutions $\Xi_{(-\infty,L)}(x,y)$ that are non-increasing in $x$ and $L$, in analogy to Lemma \ref{H_infty_L_problem_solution}. 
Exploiting these monotonicity properties, we let $L\to\infty$, suitably adapting the arguments from Section \ref{section:H_infty}. The main new aspect of the proof here arises when showing that the limits do not vanish, that is, when constructing a subsolution, in analogy to Proposition \ref{H_infty_L_asymptotics_to_profiles}.
For this, the main ingredients are contained in the following lemmas. Recall the definition of $\bar{u}(y)$ as part of a periodic solution to  $u''+u-u^3=0$. As such, $\bar{u}$ is embedded into a family of periodic solutions. 

\begin{Lemma}[Stretched periodic orbits]\label{Thm:stretched}
There exists a family of $\frac{2 \kappa}{\beta}$ periodic solutions $\bar{u}^{\beta}(\cdot) $, $\beta\geq  1$,  satisfying
$
\bar{u}^{\beta}(0) = \bar{u}^{\beta}\left(\frac{\kappa}{\beta}\right) =0$  and $0 <\bar{u}^{\beta}(y) <1$ { for} $y \in \left(0,\frac{\kappa}{\beta}\right)
.$ The map $\beta\mapsto \bar{u}^\beta\left(\frac{x}{\beta}\right)$ is continuous, uniformly in $x$. 
Furthermore, for any $\beta >1$, we have
\begin{eqnarray}\label{ordering}
\bar{u}(y) - \bar{u}^{\beta}\left(\frac{y}{\beta}\right) >0  \quad \mbox{for} \quad  y \in (0, \kappa). 
\end{eqnarray}
\end{Lemma}
\begin{Proof}
%
%
%
Phase plane analysis, exploiting the conserved quantity $H^-$ from Section \ref{s:pp}, immediately gives the existence of a family of periodic orbits parameterized by the amplitude $a$. Inspecting the formula for the period, \cite[\S V]{Hale}, 
\begin{equation}\label{period_of_periodic_orbits}
 \mathscr{K}(a) = 2 \sqrt{2} \int_{0}^1 \frac{dv}{\sqrt{[(1 - v^2)(2 - a^2(1 + v^2))]}},
\end{equation}
we see that $\mathscr{K}'>0$, such that we can use the implicit function theorem to parameterize periodic orbits by the period instead of the amplitude. Continuity of the flow gives the continuous dependence in the uniform topology on the amplitude, hence on the period.  It remains to establish the ordering relation \eqref{ordering}. 
 Write 
 %
$ v\mapsto Q^{(\beta)}[ v](\cdot):= \frac{1}{\beta^2}\partial_y^2 v(\cdot) +  v(\cdot) -  v^3(\cdot),$
and define $v^{\beta}(\cdot) := u^{\beta}\left(\frac{\cdot}{\beta} \right).$ Clearly, $ Q^{(\beta)}[\bar{v}^{\beta}](\cdot) =0.$ Furthermore, for any $\beta >1$, we have that 
 \begin{eqnarray}\label{monotonicicy_for_Q_operator}
 Q^{(\beta)}[v](\cdot) =Q^{(1)}[v](\cdot) + \frac{(1 - \beta^2)}{\beta^2}\partial_y^2 v(\cdot).
 \end{eqnarray}
 Pluging $\bar{u}(\cdot)$ in \eqref{monotonicicy_for_Q_operator} we obtain
 \begin{align*}
 Q^{(\beta)}[\bar{u}](\cdot) &=Q^{(1)}[\bar{u}](\cdot) + \frac{(1 - \beta^2)}{\beta^2}\partial_y^2 \bar{u}(\cdot) =  \frac{1 - \beta^2}{\beta^2}\partial_y^2 \bar{u}(\cdot) =  \frac{1 - \beta^2}{\beta^2}\bar{u}\left(\bar{u}^2 -1 \right) \leq 0 = Q^{(\beta)}[v^{\beta}]
 \end{align*}
The strong maximum principle applied to the elliptic operator  $Q^{\beta}[\cdot]$ now implies that $\bar{u}(y)-v^{\beta}(y) \geq 0$ for  $y \in (0,\kappa).$ This concludes the proof.


\end{Proof}

\begin{Lemma}[Sub- and supersolutions]\label{l:sub}

 Let $\alpha,\beta > 0$ such that $\frac{1}{\alpha^2} + \frac{1}{\beta^2} =1$ and define
 \begin{equation}\label{e:subs}
 w_{(\alpha, \beta)}(x,y) := \theta_{(-\infty,L)}\left(\frac{x}{\alpha}\right)\cdot \bar{u}^{\beta}\left(\frac{y}{\beta}\right)
 ,\end{equation}
where $\bar{u}^{\beta}$ was defined in Lemma \ref{Thm:stretched}. 
Then 
\begin{equation}\label{e:dir}
\Delta w_{(\alpha, \beta)} + \mu(x) w_{(\alpha,\beta)} - w_{(\alpha,\beta)}^3 \geq 0,
\end{equation}
in the sense of distributions on $\mathcal{S}_{(-\infty,\alpha\cdot L)}$ (and, consequently, in $\mathcal{S}_{(-\infty,L)} \subset \mathcal{S}_{(-\infty,\alpha\cdot L)}$). 

Now let $\widetilde{L} \leq L$,  $\alpha = L/\widetilde{L}$ and $\beta = L/\sqrt{ L^2 -\widetilde{L}^2 }.$  Then 
\begin{equation}\label{e:comp}
w_{(\alpha, \beta)}(x,y) \leq \Xi_{(-\infty,L)}(x,y) \quad \mbox{for} \quad \mathcal{S}_{(-\infty,L)}.
\end{equation}
%
We also have supersolutions $\theta(x)$ and $\bar{u}(y)$, such that 
\begin{eqnarray}\label{Thm:stretched:2D_stripe_supersolutions_infty_L}
 \Xi(x,y)\leq \min\{\theta(x), \bar{u}(y)\}. 
 \end{eqnarray}  
\end{Lemma}

\begin{Proof}
The last assertion on supersolutions follows readily from the construction of solutions using the iteration scheme. 
In order to establish \eqref{e:subs}, we follow the direct calculation of Lemma \ref{Thm:subsolutions_alpha_beta:part_1}. It remains to show  \eqref{e:comp}. For this, it is sufficient to establish the inequality for the truncated solutions, exploiting the monotonicity of the sequence of solutions and the extension operator. 
\begin{eqnarray}\label{simpler_result:H_problem}
w(x,y) \leq \Xi_{(-M,L)}(x,y)  \quad \mbox{for}  \quad (x,y) \in  \mathcal{S}_{(-M,L)}.
\end{eqnarray}
For this, i is sufficient to show that  $w$ is a subsolution to  \ref{2D_H_infty_problem} in $\mathcal{S}_{(-M,L)}$ since then the iteration process will stay strictly above $w$. Since $w$ is a subsolution in the interior of the domain, \eqref{e:dir}, we only need to check boundary values. 
On $x \in (-M,L)$, $y = 0,$
$$[\Xi_{(-M,L)} - w](x,0) = h_{(-M,L)}(x,0) - \theta_{(-M,L)}(x)\bar{u}^{\beta}(0) = 0.$$
 On $x \in (-M,L)$, $y = \kappa,$
 \begin{align*}
  [\Xi_{(-M,L)} - w](x,0) = h_{(-M,L)}(x,\kappa) - \theta_{(-M,L)}(x)\bar{u}^{\beta}\left(\frac{\kappa}{\beta}\right) = 0.
 \end{align*}
 On $x = L$, $y \in (0,\kappa),$ 
 \begin{align*}
[\Xi_{(-M,L)} - w](L,y)  &=  h_{(-M,L)}(L,y) - \theta_{(-M,\widetilde{L})}\left(\frac{L \cdot \widetilde{L}}{L}\right)\bar{u}^{\beta}\left(\frac{y}{\beta} \right)  \\
&  = \theta_{(-M,L)}(L)\bar{u}(y) - \theta_{(-M,\widetilde{L})}(\widetilde{L})\bar{u}^{\beta}\left(\frac{y}{\beta} \right)  =0.
 \end{align*}
On $x = -M$, $y \in (0,\kappa),$
 \begin{align*}
[\Xi_{(-M,L)} - w](-M,y)  &=  h_{(-M,L)}(-M,y) - \theta_{(-M,\widetilde{L})}\left(\frac{-M}{\alpha}\right) \bar{u}^{\beta}\left(\frac{y}{\beta}\right)  \\
&  = \theta_{(-M,L)}(-M)\bar{u}(y) - \theta_{(-M,\widetilde{L})}\left(\frac{-M}{\alpha}\right) \bar{u}^{\beta}\left(\frac{y}{\beta}\right)  \\
&  \geq \bar{u}(y) - \bar{u}^{\beta}\left(\frac{y}{\beta}\right){\geq} 0,
 \end{align*}
where in the last inequality we used Lemma \ref{Thm:stretched}. Summarizing, we have 
$\left(h_{(-M,L)}(x,y)  - w(x,y)\right)\big|_{\partial \mathcal{S}_{(-M,L)} }\geq 0$, which together with \eqref{e:dir} implies  $w(x,y)\leq \Xi_{(-M,L)}(x,y)$ in $\mathcal{S}_{(-M,L)}.$ Finally, one readily sees that $\theta(x)$ is a supersolution, taking strictly positive boundary values, and $\bar{u}(y)$ is a supersolution since the residual of the equation is negative in $x>0$, which proves the lemma. 
 \end{Proof}
Using this result, one readily establishes the uniform convergence of solution $\Xi_{(-\infty,L)}$  to the limit $\bar{u}(y)$ as $x\to -\infty$; see the proof of Proposition \ref{H_infty_L_asymptotics_to_profiles}. Moreover, using again monotonicity in $L$, we can let $L\to\infty$ and find the existence of a solution to the $\mathcal{H}$-problem. Exploiting the fact that the subsolution $w_{(\alpha,\beta)}(x,y)$ converges to $\bar{u}^\beta(\frac{y}{\beta})$ for $x\to -\infty$, and the fact that $\bar{u}^\beta(\frac{y}{\beta})\to\bar{u}(y)$, uniformly in $y$ as $\beta \to 1$, we establish convergence of $\Xi(x,y)$ to $\bar{u}(y)$ as $x\to-\infty$.

We omit details that are analogous to but slightly simpler than in the $\mathcal{H}_\infty$-case.

%


\section{Non-zero speed --- existence and non-existence}\label{section:c_greater_zero:Allen_Cahn}

We prove results for non-zero speeds in the Allen-Cahn, Section \ref{s:acc}, and in the Cahn-Hilliard setting, Section \ref{s:chc}, perturbing from the profiles we found in the previous sections for the zero-speed case. 

\subsection{Existence --- Allen-Cahn}\label{s:acc}
Our goal is to prove Proposition \ref{t:acc}. We start with the case $\kappa<\infty$, $\mathcal{H}$-problem, and then present the necessary modifications in the case $\kappa=\infty$, $\mathcal{H}_\infty$-problem. The proof is based on a study of the linearization at the solutions for $c=0$. We wish to study 
\begin{align}\label{Allen_Cahn_parametrized_problem}
\Delta u + \mu(x) u  - u^3  + c u_x =0, \quad (x,y)\in \R\times (0,\kappa),
\end{align}
near $c=0$ and $u=\Xi(x,y)$, the solution from Theorem \ref{main_theorem:c_zero:H_infty} at $\kappa<\infty$. Linearizing at this solution, we find the elliptic operator 
\begin{align*}
   \mathscr{L}^{AC}_{\Xi}[v] :=  \Delta v + \left[\mu(x)  - 3 \Xi^2(x,y)\right]v   + c v_x,
\end{align*}
 with domain of definition $ \mathcal{D}\left(\mathscr{L}_{\Xi}^{AC}\right) = H^2\left(\mathbb{R}\times [0,\kappa]\right) \cap H_0^1\left(\mathbb{R}\times [0,\kappa]\right)$.

\paragraph{The linearization, $\kappa<\infty$.} We show that $\mathscr{L}^{AC}_{\Xi}$ is bounded invertible. We therefore show first that the essential spectrum of $\mathscr{L}^{AC}_{\Xi}$ is strictly negative, implying that  $\mathscr{L}^{AC}_{\Xi}$ is Fredholm of index 0. We then show that any element in the kernel would be exponentially localized and continue to show that the kernel is trivial since the point spectrum is negative. 

To get started, we study the limiting operators in the far field, $x=\pm\infty$. 
\begin{Lemma}[Far-field operators]\label{spectrum_of_M:H_problem} Let  $\bar{u}(\cdot)$  be a solution to 
$$\partial_y^2\bar{u}  + \bar{u}- \bar{u}^3=0,\quad \bar{u}(y) >0, \quad \mbox{for} \quad y \in (0, \kappa), \quad \mbox{and}\quad u(0)= u(\kappa) =0 .$$
Then the operators 
 \begin{align*}
\mathscr{M}^-[v] := &\partial_y^2v(\cdot)  + (1- 3\bar{u}^2(y))v(\cdot),  \\
\mathscr{M}^+[v] := & \partial_y^2v(\cdot)  -v(\cdot),        
\end{align*}
with domain             $\mathcal{D}\left(\mathscr{M}^+ \right) =  H^2([0,\kappa])\cap H_0^1([0,\kappa])$ are self-adjoint negative definite, that is, their spectra  satisfy $\sigma(\mathscr{M}^-)<0$ and $\sigma(\mathscr{M}^+)<0$.
\end{Lemma}
 
\begin{Proof}
We prove the result for $\mathscr{M}^{-}$ only; the analysis for $\mathscr{M}^{+}$ is easier. We first note that the operator $\mathscr{M}^-[\cdot]$ is self-adjoint, bounded from above (see for instance  \cite{kato2013perturbation}[V-\S 4]) and thereby  $\sigma\left(\mathscr{M}^-\right) \subset (-\infty,M)$, for some $M<\infty$. Since the domain is compactly embedded into $L^2$, the spectrum consists entirely of point spectrum.  We also know that the largest eigenvalue $\lambda \in \sigma(\mathscr{M}^-)$ is simple and its associated eigenfunction $v$ has no nodal points, i.e., we can take $v(\cdot) >0$  in $y \in (0,\kappa)$ (cf. \cite[Thm. 8.38]{gilbarg2015elliptic}). The result follows if we show that $\lambda <0$. Assume that $\lambda \geq 0$. We have
 \begin{eqnarray*}
v'' + v - 3 \bar{u}^2 v = \lambda v,\qquad 
 \bar{u}''  + \bar{u}- \bar{u}^2 \bar{u}= 0.
 \end{eqnarray*}
We multiply the first equation by $\bar{u}$ and the second equation by $v$, subtract and integrate over $[0,\kappa]$, to obtain
 $$\int_0^{\kappa}\left\{ \left(v''\bar{u} - \bar{u}''v\right) -2\bar{u}^2\bar{u}v\right\}  dx= \lambda \int_0^{\kappa} \bar{u}v \geq 0.$$
 Integrating by parts on left hand side and using the boundary conditions we find
 $$\int_0^{\kappa}\left\{ \left(v''\bar{u} - \bar{u}''v\right) -2\bar{u}^2\bar{u}v\right\}  dx=\int_0^{\kappa} -2\bar{u}^2\bar{u}v  dx <0,$$
 contradicting the fact that $\bar{u}^3v$ is strictly positive.
 \end{Proof}
 We now readily deduce Fredholm properties of the linearization $\mathscr{L}^{AC}_{\Xi}$.
 \begin{Lemma}\label{Allen_Cahn-c_greater_zero-essential_spectrum:H_problem} The operator $\mathscr{L}^{AC}_{\Xi}$ is self-adjoint, Fredholm of index zero, with essential spectrum strictly negative.
 \end{Lemma}
\begin{Proof}
We consider $\mathscr{L}^{AC}_{\Xi}-\lambda$ for $\lambda\geq 0$. Note that this operator is invertible for all $\lambda\gg 1$ since $\mathscr{L}^{AC}_{\Xi}$ is a bounded perturbation of the Laplacian.  We claim that in addition $\mathscr{L}^{AC}_{\Xi}-\lambda$ is Fredholm for all $\lambda\geq 0$, which then implies the proposition, by continuity of the Fredholm index. The fact that $\mathscr{L}^{AC}_{\Xi}-\lambda$ is Fredholm of index zero follows immediately from the Closed Range Lemma, exploiting that the limiting operators are invertible and that $\mathscr{L}^{AC}_{\Xi}$ is self-adjoint; see for instance \cite[\S 3]{rs}.
\end{Proof}
We next eliminate the possibility of a non-trivial kernel. We therefore first show that elements of the kernel are exponentially localized in the $x$-direction.
\begin{Lemma}[Exponential decay of eigenfunctions] Let $\lambda-\mathscr{L}_{\Xi}^{AC}$ be Fredholm of index 0 and $u$ a nontrivial element of the kernel. Then there exist  $C,\delta >0$ such that 
\begin{equation*}
|\nabla u(x,y)|+|u(x,y )| \leq  C e^{- \delta|x|} ,\quad  (x, y)  \in  \mathbb{R}\times [0,\kappa] \quad a.e.
\end{equation*}
\end{Lemma}
\begin{Proof}
The analysis is similar to those in \cite[Proposition 2.1]{Jaramillo_Scheel_2015} and \cite[Lemma 5.7]{sandstede_scheel_2001structure} and relies solely on the Fredholm properties. Assume $\lambda=0$ without loss of generality. The functions $v(x,y)=e^{\delta\langle x\rangle}u(x,y)$ belong to the kernel of $\mathscr{L}^{AC}_{\Xi,\delta}:=e^{\delta\langle x\rangle}\mathscr{L}^{AC}_{\Xi}e^{\delta\langle x\rangle}$, with domain of definition independent of $\delta$. A direct computation shows that $\mathscr{L}^{AC}_{\Xi,\delta}$ is an analytic family of operators in the parameter $\eta$, and therefore Fredholm of index 0 for all $\delta$ small. From the construction,  $\mathrm{Ker}\left(\mathscr{L}_{\Xi,\delta}^{AC}\right)\subset\mathrm{Ker}\left(\mathscr{L}_{\Xi,\delta'}^{AC}\right)$, for $\delta'<\delta$, which implies  $\mathrm{Ker}\left(\mathscr{L}_{\Xi,\delta}^{AC}\right)\equiv\mathrm{Ker}\left(\mathscr{L}_{\Xi,\delta'}^{AC}\right)$, by analyticity; see for instance \cite[Sections I-$\S$4.6 and III-$\S$3.4]
{kato2013perturbation} . Hence, $v(x,y)\in H^2$, which, using Sobolev embeddings, gives the exponential  decay of $u(\cdot+L,\cdot)$ in $H^2([0,1]\times[0,\kappa])$ in $L$. A bootstrap now shows that in fact $u(\cdot+L,\cdot)$  decays exponentially in $\mathscr{C}^{1,\alpha}([0,1]\times[0,\kappa])$, implying in particular the decay estimate as stated. 
\end{Proof}

\begin{Lemma}\label{Allen-Cahn-negative_spectrum:H_problem}\label{L_is_invertible:Allen_Cahn:c_greater_zero:H_problem} The spectrum of the operator $\mathscr{L}_{\Xi}^{AC}$ satisfies $\sigma\left(\mathscr{L}_{\Xi}^{AC}\right)< 0.$ In particular, the operator $\mathscr{L}_{\Xi}^{AC}$ is invertible.
\end{Lemma}
\begin{Proof}
Recall that eigenvalues are real valued. Integration by parts and boundedness of the operator $u \mapsto \left(\mu(x) - 3 \Xi^2\right)u$ implies that the eigenvalues of the operator $\mathscr{L}_{\Xi}^{AC}$ are in a bounded set of the real line. 
%
%
Let $\lambda_0 \in \mathbb{R}$ be the largest eigenvalue in the spectrum of $\sigma\left(\mathscr{L}_{\Xi}^{AC} \right).$ The result will follow if we show that $\lambda_0 < 0$. Arguing by contradiction, assume that $\lambda_0 \geq0$ with associated eigenfunction $u_0$, that is,  
\begin{equation}\label{eigenvalue_not_positive_eq1}
\mathscr{L}_{\Xi}^{AC}u_0 = \Delta u_0  + (\mu(x) - 3\Xi^2)u_0 = \lambda_0 u_0 
\end{equation}
Recall that $ \Delta\Xi  + \mu(x)\Xi - \Xi^3 = 0.$ Multiply this equation by $\Xi$ and \eqref{eigenvalue_not_positive_eq1} by $u_0$, subtract both and integrate to find 
\begin{equation*}
\int_{\mathbb{R}\times[0,\kappa]} (\Xi \Delta u_0 -u_0  \Delta \Xi)dxdy - 2\int_{\mathbb{R}\times[0,\kappa]}\Xi^3 u_0 dx dy = \lambda_0 \int_{\mathbb{R}\times[0,\kappa]} u_0 \Xi dx dy.
\end{equation*}
Since $u_0$ corresponds to the eigenfunction associated to the maximal element in the spectrum of an elliptic self-adjoint operator, we can assume that $u_0 \geq 0$ a.e. (cf. \cite[Thm. 8.38]{gilbarg2015elliptic}). We integrate in the whole space and use the exponential decay as $|x| \to \infty$ to obtain
\begin{equation*}
 -2 \int_{\mathbb{R}\times[0,\kappa]}\Xi^3 u_0 dx dy = \lambda_0 \int_{\mathbb{R}\times[0,\kappa]} u_0 \Xi dx dy.
\end{equation*}
Since $\Xi > 0$, the above equality has a non-negative left hand side but a non-positive right hand side. We conclude that the left hand side is zero, which implies that $u_0 \equiv 0$ a.e., contradicting positivity of the first eigenfunction and proving the result. 
\end{Proof}

\begin{Corollary}[Exponential convergence $\mathcal{H}$]\label{c:exp}
The convergence $\lim_{x\to-\infty}\Xi(x,y)=\bar{u}(y)$, $\lim_{x\to\infty}\Xi(x,y)=0$, is exponential with differences bounded by $Ce^{-\delta|x|}$ for some $C,\delta>0$, uniformly in $y$. 
\end{Corollary}
\begin{Proof}
Let $\chi$ be smooth cut-off function with $\chi(x)=0$ for  $|x|\leq 1$ and  $\chi(x)=1$ for$|x|\geq 2$. Multiplying \eqref{Allen_Cahn_parametrized_problem} by $\chi$ and differentiating, one quickly computes that $v=(\chi \Xi)_x$ satisfies an equation 
$\mathscr{L}^{AC}_{\Xi} v = f$, where $f$ is compactly supported, smooth. Using the fact that  $\mathscr{L}^{AC}_{\Xi} $ is bounded invertible in exponentially weighted spaces $e^{\delta \langle x\rangle}v\in H^2\cap H^1_0$, we find exponential decay of $\Xi_x$ and hence of $\Xi$ from Sobolev embedding.  
\end{Proof}

We finish this section with the proof of our main result of this section.
 
\paragraph{Nonlinear perturbation via IFT.}
We prove Proposition \ref{t:acc} in the case $k>0$ ($\mathcal{H}$). 
We rewrite  \eqref{Allen-Cahn_c_greater_than_zero} as
 \begin{align}\label{Linear_equals_nonlinear:c_greater_zero:Allen_Cahn:H_problem}
  \mathscr{L}^{AC}_{\Xi}[v] = \mathscr{N}^{AC}_{\Xi}[v;c] := -c(v + \Xi)_x  + 3\Xi v^2 + v^3.
 \end{align}
 We consider this as an equation for $v$ with parameter $c$, mapping $v\in (H^2\cap H^1_0)(\mathbb{R}\times[0,\kappa])$ into $L^2$. 
 In order to apply the implicit function theorem near $c=0$ and $v=0$, we only need to show that the nonlinearity $\mathscr{N}^{AC}_{\Xi}[v] $ is $C^1$ in $v$ and $c$. For this, first note that the superposition operators $v\mapsto v^2$ and $v\mapsto v^3$ as maps from $(H^2\cap H^1_0)(\mathbb{R}\times[0,\kappa])$ into itself are smooth, owed to the embedding into $L^\infty$, which makes $H^2$ an algebra. Moreover, multiplication by $\Xi$ is bounded from $H^2$ into $L^2$ and $-cv_x$ defines a bounded linear operator. it remains to show $\Xi_x\in L^2$. Since $\Xi_x$ is exponentially decaying and H\"older continuous, hence bounded, we have $\Xi_x\in L^2$, which proves that $\mathscr{N}^{AC}_{\Xi}$ is a smooth map, and we can find the family of solutions parameterized by $c$ as stated in Proposition \ref{t:acc}.

\paragraph{Pertubing the $1\leadsto 0$-problem.}

The analysis performed in the previous section applies here without major modifications and we can conclude from the Implicit Function Theorem the claim. on the existence of profiles. In this particular case, one could also pursue the geometric proof of existence, Section \ref{s:pp}, and find intersections of the now $c$-dependent stable and unstable manifolds. 
  
%

\paragraph{Perturbing the $\mathcal{H}_\infty$-problem.}

Let $\Theta$ be the solution to the $\mathcal{H}_{\infty}$-problem found in Section \ref{section:H_infty}. We linearize \eqref{Allen_Cahn_parametrized_problem} at $(u, c) = (\Theta,,0)$, and obtain the linear operator
\begin{align*}
  \mathscr{L}^{AC}_{\Theta}[v] &:=  \Delta v + \left[\mu(x)  - 3 \Theta^2(x,y)\right]v, 
\end{align*}
with domain of definition $\mathcal{D}\left(\mathscr{L}_{\Theta}^{AC}\right) = (H^2\cap H_0^1)(\mathbb{R}\times (-\infty, 0)).$

Roughly speaking, at this point we'd like to pursue the strategy from the previous section. A difficulty that arises is that the $c$-dependent correction at $y=\infty$ is non-trivial, such that at a fixed finite $x$-value, the correction to $\Theta$ will not belong to $L^2$. There are at least two remedies. One could choose to work in spaces where the correction to the profile $\Theta$ lies in spaces of bounded functions, for instance uniformly local spaces. Alternatively, one could explicitly incorporate the corrections at $y=\pm\infty$ into the ansatz and solve for a remaining, localized correction. We will pursue this later strategy after analyzing the linearization. 
We start with the lienarization. 

\begin{Lemma}\label{Allen_Cahn-c_greater_zero-essential_spectrum:H_inftyproblem} The operator $\mathscr{L}^{AC}_{\Theta}$ is self-adjoint, Fredholm of index zero, with essential spectrum strictly negative.
\end{Lemma}
\begin{Proof}
The proof is identical to the proof of Lemma \ref{Allen-Cahn-negative_spectrum:H_problem} and will be omitted here.
\end{Proof}

\begin{Corollary}[Exponential convergence $\mathcal{H}_\infty$]\label{c:expinf}
The convergence 
\[
\lim_{x\to\infty}\Theta(x,y)=\tanh(y/\sqrt{2}),\qquad \lim_{y\to\infty}\Theta(x,y)=\theta(x),\qquad  \lim_{x\to-\infty}\Theta(x,y)=0,
\]
is exponential with differences bounded by $Ce^{-\delta|x|}$  and $Ce^{-\delta|y|}$f or some $C,\delta>0$, uniformly in $y$ and $x$, respectively. 
\end{Corollary}
As a next step, we  construct a suitable ansatz with farfield corrections that will allow us to apply the implicit function theroem.
%
%
\begin{Proof}
The proof is analogous to the proof of Corollary \ref{c:exp}. With the notation from there, we find exponential localization of $\chi \Theta_x$, establishing convergence as $|x|\to\infty$. On the other hand, using a smooth cut-off function $\chi$ with $\chi=1$ for $|y|>2$, $\chi=0$ for $|y|<1$, we see that $\chi \Theta_y$ solves an equation with exponentially localized right-hand side, therefore is exponentially localized, thus establishing the exponential convergence in $y$.
\end{Proof}
We set $u(x,y)=\Theta(x,y)+\chi_1(y)(\theta(x;c)-\theta(x;0))+v(x,y)$, with $\chi_1(y)$ smooth, $\chi_1(y)=0$ for $\chi<1$, $\chi_1(y)=1$ for $y>2$. Substituting this ansatz into the equation, we find, writing $\theta^c(\cdot)$ for $\theta(\cdot;c)$,
\begin{equation}\label{Linear_equals_nonlinear:c_greater_zero:Allen_Cahn:H_infty_problem}
\mathscr{L}^{AC}_{\Theta}[v]+ \mathscr{N}^{AC}_{\Theta}[v;c]=0,
\end{equation}
 where a short calculation gives
\begin{align*}
\mathscr{N}^{AC}_{\Theta}[v;c]=&[\Delta,\chi_1](\theta^c-\theta^0)+c(\Theta_x-\chi_1\theta_x^0)+cv_x\\
&-\left(\chi_1^3-\chi_1\right)\left((\theta^c)^3-(\theta^0)^3\right) -3(\Theta+(\theta^c-\theta^0)\chi_1)v^2-v^3.
\end{align*}
Here, the commutators of linear part $[\Delta,\chi_1]$ and nonlinear part $\chi_1^3-\chi_1$ generate $v$-independent terms with compact support. Also, $\Theta_x-\chi_1\theta^0$ is exponentially localized by Corollary \ref{c:expinf}. Since linear and nonlinear terms in $\mathscr{N}^{AC}$ are smooth on $H^2$, we conclude that  $\mathscr{N}^{AC}$ is well-defined and smooth. The implicit function theorem then gives the desired result.

\subsection{Existence --- Cahn-Hilliard}\label{s:chc}
Our approach for Cahn-Hilliard is slightly different from what we did for Allen-Cahn, $c >0$, and we discuss it formally here before starting. Rather than attempting to analyze the full fourth order elliptic linearization at a solution $\Xi$, we rewrite \eqref{Cahn-Hilliard_c_greater_than_zero}  as
\begin{equation}\label{Cahn_Hilliard_parametrized_problem}
  -c\left(\Delta\right)^{-1} u_x  = \left[\Delta u + \mu(x) u  - u^3\right],
\end{equation}
assuming of course that the Laplacian is in fact invertible. 

\begin{Lemma}[Invertibility of the Laplacian]\label{l:delinv}
 The operator $\Delta$  on $L^2(\mathbb{R}\times [0, \kappa])$ with domain of definition $\mathcal{D}\left(\Delta\right) = (H^2\cap H^1_0)(\mathbb{R}\times [0, \kappa]) $ 
 is bounded invertible.
\end{Lemma}
\begin{Proof}
After Fourier transform in $x$, the operator is conjugate to $\partial_{yy}-k^2$, $k\in\R$. Since $\partial_{yy}$, equipped with Dirichlet boundary conditions, is negative definite, we find an inverse that is uniformly bounded in $k$, establishing boundedness of the inverse in physical space. 
\end{Proof}
Now let $\Xi$ be the solution to the $\mathcal{H}$problem found in Section \ref{section:H_problem}. We linearize  \eqref{Cahn_Hilliard_parametrized_problem} at $(u, c) = (\Xi,0)$, obtaining
\begin{align*}
  \mathscr{L}^{CH}_{\Xi}[v] :=  \Delta v + \left[\mu(x)  - 3 \Xi^2(x,y)\right]v, 
\end{align*}
with domain $\mathcal{D}\left(\mathscr{L}_{\Xi}^{CH}\right)= (H^2\cap H^1_0)\left(\mathbb{R}\times [0,\kappa]\right)$.
We know from Lemma \ref{L_is_invertible:Allen_Cahn:c_greater_zero:H_problem} that $\mathscr{L}^{CH}_{\Xi}[v] $ is bounded invertible from $L^2\left(\mathbb{R}\times [0,\kappa]\right)$ to $\mathcal{D}\left(\mathscr{L}_{\Xi}^{CH}\right)$. We rewrite problem \ref{Allen_Cahn_parametrized_problem} as
 \begin{align}
  \mathscr{L}^{CH}_{\Xi}[v] -\mathscr{N}^{CH}_{\Xi}[v;c] =0,\qquad  \mathscr{N}^{CH}_{\Xi}[v;c] := \left(\Delta \right)^{-1}\left[-c(v + \Xi)_x  + 3\Xi v^2 + v^3\right].
 \end{align}
We now readily verify as in the proof of Theorem \ref{t:acc} that $\mathscr{N}_{\Theta}^{CH}[v,c]$ is smooth as a map from $\mathcal{D}\left(\mathscr{L}_{\Xi}^{CH}\right)$ into $L^2 $. We already showed that all $c$-independent terms are smooth in the Allen-Cahn case.  By Lemma \ref{l:delinv}, $\Delta ^{-1}$ is bounded from $L^2$ to $H^2$, which completes the proof of smoothness of $\mathscr{N}_{\Theta}^{CH}$ and completes the proof of Proposition \ref{t:chc}.

 \section{Discussion and open problems}\label{open_problems}
We discuss some open questions arising from the present work. First, one can clearly generalize in many ways, allowing for more general jump discontinuities $f(x,u)=f_\pm(u)$, $\pm x>0$, with $f_+(u)u>0$ and $f_-(u)=-f_-(-u)$ bistable, or even more general homotopies from monotone to bistable nonlinearities as $x$ varies from $+\infty$ to $-\infty$. We expect many of our methods to allow for such extensions. Possibly somewhat restrictive is our assumption on monotonicity of the period of periodic solutions in the amplitude of periodic patterns. 

\paragraph{Selection of an angle.}
We view the present work as a first attempt to gain a better understanding of selection processes induced by directional quenching. It appears that the dynamic process, $c>0$, selects patterns to some extent: given a latteral $y$-period, one expects an angle selection. We showed here, that this angle tends to be $\pi/2$, that is, stripes align perpendicular to the interface. Recent work on the Swift-Hohenberg equation \cite{GS} shows that one expects a family of oblique stripes with $\phi\sim 0$ to accompany a time-periodic solution forming vertical stripes, that is existence of solutions as depicted in Figure \ref{Figure_1-horizontal}, top right. Given the results in \cite{krekhov2009formation}, one would therefore expect oblique stripes in Cahn-Hilliard as well. It is an interesting question to explore the angle in this family of oblique stripes as the lateral period is decreased.
 
\paragraph{Stability.}
We did not perform a stability analysis of these patterns. In the Allen-Cahn case, stripes are unstable and any stability result would need to be understood in a pointwise sense. One could for instance focus on the absence of singularities of the Green's function, that is, the absence of zeros of an analytic extension of the Evans function, first. On the other hand, restricting to the reduced domains, with Dirichlet boundary conditions at $y=0$ and $y=\kappa$, we would expect the solutions found here to be stable. In fact, interpreting our iteration schemes used in the proof as an implicit Euler time stepping for the parabolic flow, we recognize that, at least in the truncated problems, stability is a prerequisite for showing existence using our methods. An interesting marginal case is the $\mathcal{H}_\infty$ case, where a single interface is created. We suspect that this solution $\Theta$ is stable also as a solution in $\R^2$, although decay would be slow as the relaxation of an infinitely bent interface 
happens on a diffusive time scale.  

\paragraph{Other nonlinearities.}
As pointed out before, our analysis should apply to a large class of odd nonlinearities. Dropping this symmetry, the problem is however wide open. One can readily suspect that for non-balanced nonlinearities, $\int_{u_-}^{u_+}f(u)\neq 0$, where $u_\pm$ are the two stable equilibria, interfaces created at the trigger would move, but establishing existence and selection laws for the angle, for single interfaces or periodic interface arrangements behind the trigger, appears to be challenging. 

\paragraph{Spatial dynamics and completeness.}
We intend to pursue a more detailed analysis of the existence and stability problem for $\kappa\gtrsim \pi$, studying the solutions near $u=0$ using spatial $x$-dynamics and center-manifold reduction. The hope is to reduce the existence problem to a geometric picture similar to the one-dimensional problem studied in Section \ref{s:pp}. This would allow for a more systematic study of angle selection, in particular also allowing for Cahn-Hilliard dynamics, at least in this limit of small lateral period.

\fontsize{10}{10}\selectfont

\end{document}